\documentclass[apj]{emulateapj}
\usepackage{apjfonts}

\begin{document}


\title{The Structure of Classical Bulges And Pseudobulges: The Link
  Between Pseudobulges And S\'ersic Index}

\shorttitle{Structure of Pseudobulges \& Classical Bulges}
\shortauthors{Fisher \& Drory}


\author{David~B.~Fisher\footnote{For complete version with appendix and high resolution images, please visit http://spitzer.as.utexas.edu/~twitch/Papers/papers.html}}
\affil{Department of Astronomy, The University of Texas at Austin,\\
  1 University Station C1400, Austin, Texas 78712\\
  {\tt dbfisher@astro.as.utexas.edu}}

\and

\author{Niv~Drory}
\affil{Max-Planck-Institut f\"ur
  Extraterrestrische Physik, Giessenbachstra\ss e, 85748 Garching, Germany\\
  {\tt drory@mpe.mpg.de}}

\slugcomment{Submitted to ApJ}


\begin{abstract}
  In this paper we study the properties of pseudobulges (bulges that
  appear similar to disk galaxies) and classical bulges (bulges which
  appear similar to E-type galaxies) in bulge-disk decompositions. We
  show that the distribution of bulge S\'ersic indices, $n_b$, is
  bimodal, and this bimodality correlates with the morphology of the
  bulge.  Pseudobulges have $n_b \lesssim 2$ and classical bulges have
  $n_b \gtrsim 2$ with little-to-no overlap.  Also, pseudobulges do
  not follow the correlations of S\'ersic index with structural
  parameters or the photometric projections of the fundamental plane
  in the same way as classical bulges and elliptical galaxies do.  We
  find that pseudobulges are systematically flatter than classical
  bulges and thus more disk-like in both their morphology and
  shape. We do not find significant differences between different
  bulge morphologies that we are collectively calling pseudobulges
  (nuclear spirals, nuclear rings, nuclear bars, and nuclear
  patchiness) appear to behave similarly in all parameter
  correlations. In S\'ersic index, flattening, and bulge-to-total
  ratio, the distinction appears to be between classical bulges and
  pseudobulges, not between different pseudobulge morphologies. The
  S\'ersic index of the pseudobulges does not correlate with $B/T$, in
  contrast to classical bulges.  Also, the half-light radius of the
  pseudobulge correlates with the scale length of the disk; this is
  not the case for classical bulges.  The correlation of S\'ersic
  index and scale lengths with bulge morphology suggests that secular
  evolution is creating pseudobulges with low-S\'ersic index, and that
  other processes (e.g.~major mergers) are responsible for the higher
  S\'ersic index in classical bulges and elliptical galaxies.
\end{abstract}



\section{Introduction}\label{sec:introduction}

Historically, all bulges were assumed to be little elliptical galaxies
residing in the centers of galactic disks. Thus it was also assumed
that all bulges were dynamically hot stellar systems. Recent work puts
this assumption in question. Many bulges have disk-like features that
do not resemble E-type galaxies.

{\em Classical bulges} are dynamically hot, and relatively
featureless; they appear similar to the end products of galaxy major
mergers.  They are easily recognized as having morphologies very
similar to E-type galaxies.

In contrast, many bulges have features that are not found in
elliptical galaxies, but in galactic disks. These features include the
following: kinematics dominated by rotation \citep{k93}; flattening
similar to that of their outer disk \citep{fathi2003,k93}; mid-IR
colors of the bulge are similar to those of the outer disk
\citep{fisher2006}; nuclear bar \citep{erwin2002}; nuclear ring and/or
nuclear spiral \citep{carollo97}; near-exponential surface brightness
profiles \citep{andredak94}. Bulges with such properties are called
{\em pseudobulges}. All of these phenomena are manifestations of
stellar systems that that are dynamically cold. However, the extent to
which these features exist simultaneously in all pseudobulges is not
yet well-understood. For a review of the properties of pseudobulges
see \cite{kk04} (KK04 hereafter).

In this paper, we refer to the super-set of the two systems simply as
{\em bulges}.  We note that in this paper the term {\em pseudobulge}
refers to bulges with morphology reminiscent of disk galaxies; there
is no {\em a priori} assumption about their formation mechanism.

Investigating the link between bulge morphology and bulge structural
parameters is the main interest of this article. More precisely, those
bulges with morphologies that are reminiscent of disks (such as
nuclear rings, nuclear bars and nuclear spirals) are expected to have
lower S\'ersic index than those bulges with smooth light distributions
resembling E-type galaxies. We will investigate whether this
expectation holds and whether the distribution of bulge S\'ersic
indices is dichotomous.

Determining whether a critical S\'ersic index that discriminates
between classical bulges and pseudobulges exists, and what its value
is, impacts our understanding of bulge formation in at least two
ways. First, the distribution of S\'ersic indices would constrain
formation theories of classical bulges and pseudobulges. Also, the
existence of a critical S\'ersic index would robustly establish a
method for distinguishing pseudobulges from classical bulges without
using high-resolution imaging or kinematic data, neither of which is
currently available in large surveys.

In the last ten years, the subject of surface brightness profiles of
bulges and elliptical galaxies has experienced a shift of
paradigm. Traditionally, surface brightness profiles of elliptical
galaxies and bulges were all thought to be of a single shape that is
well-characterized by the de Vaucauleurs $r^{1/4}$
profile. \cite{caon94} show that surface brightness profiles of
elliptical galaxies are better fit by S\'ersic profiles
\citep{sersic1968}, which generalize the exponent in the de
Vaucauleurs profile to a free parameter. Also, \cite{andredak94} show
that many bulge-disk galaxies are better described by a double
exponential than an inner $r^{1/4}$ profile with an outer
exponential. \cite{andredak95} generalize this to show that bulge-disk
galaxies are better fit by inner S\'ersic profiles with outer
exponential disks than double exponentials profiles. The S\'ersic
function plus outer disk model for bulge-disk galaxies reads
\begin{equation}
I(r)=I_0\exp\left[-(r/r_0)^{1/n_b} \right ] + I_d\exp\left[-(r/h) \right ]\, ,
\end{equation}
where $I_0$ and $r_0$ represent the central surface brightness and
scale length of the bulge, $I_d$ and $h$ represent the central surface
brightness and scale length of the outer disk, and $n_b$ represents
the bulge S\'ersic index.

It is not surprising that S\'ersic profiles fit bulge surface
brightness profiles better than $r^{1/4}$-profiles, since the S\'ersic
function has more flexibility due to the extra parameter. However, the
new parameter, $n_b$, correlates with many properties of the stellar
systems to which it is fit, including but not limited to the
following: velocity dispersion $\sigma$ \citep{photplane} , absolute
magnitude \citep{graham96}, and effective radius \citep{caon94}. Many
authors have shown that these correlations extend to bulges of
bulge-disk galaxies
\citep{graham2001,macarthur2003,dejong2004,thomas2006}. Additionally,
\cite{andredak95} show that the S\'ersic index of bulges correlates
with Hubble type, decreasing from $n_b \sim 3.7$ for S0 galaxies to
$n_b \sim 1.6$ in Sbc-Sd galaxies. Therefore it is reasonable to
assume that it has physical significance. The S\'ersic index is often
referred to as the shape parameter, as it is generally taken as a
surrogate for properties such as concentration of the surface
brightness profile. For a review of the properties of S\'ersic
profiles see \cite{graham2005}.

The tentative assumption is that the S\'ersic index of a bulge
reflects the classical bulge - pseudobulge dichotomy. Lower S\'ersic
index might indicate that a bulge is more likely to be a pseudobulge.
We do not understand the mechanism that is responsible for determining
the S\'ersic indices in pseudobulges (or classical bulges). Yet, it
seems plausible that the light distribution be similar to that of a
disk, since so many other of the properties of pseudobulges are
similar to those of galactic disks.  \cite{courteau1996} use bulge-disk
decompositions of 243 galaxies to show that the 85\% of bulges in
Sb-Sc galaxies are better fit by the double exponential than cuspier
$r^{1/4}$ models.  Thus, the common conclusion is that pseudobulges
are marked by near exponential S\'ersic index
(KK04). \cite{scarlata2004} shows with STIS acquisition images that
bulges with surface brightness profiles more resembling exponential
profiles are more likely to have disk-like morphology (e.g.~spiral
arms), yet there is significant scatter to this claim, in their
sample. They go on to show that the distribution of central slopes of
surface brightness profiles of the bulges in their sample is bimodal
when plotted against absolute magnitude.

It is yet unknown where pseudobulges should lie in other parameter
correlations, such as fundamental plane projections. We do not expect
that pseudobulges occupy a significantly different location than
classical bulges in fundamental plane parameter space, since the
fundamental plane is not known to be bimodal. Many studies of the
locations of bulges in structural parameter space exist
(e.g.~\citealp{bbf92,graham2001,macarthur2003,dejong2004,thomas2006}),
and no significant bimodal behavior is noticed.  \cite{carollo1999}
remarks, though, that pseudobulges deviate more from the $\mu_e-r_e$
relation \citep{k77}. We will investigate this further in this
paper. \cite{k93} shows that the majority of bulges are rounder than
the outer parts of the disks they reside in, yet a significant
minority are as flat as their associated outer disk. A few bulges are
even flatter than their outer disk. This behavior correlates with
Hubble type; bulges in later type galaxies have flattening more
similar to that of the associated outer disk.  \cite{fathi2003}
carried out bulge-disk decompositions of 70 galaxies on
higher-resolution data finding a similar result.

As discussed in KK04, pseudobulges are characterized principally by
having less random motion per unit stellar light. They are
rotation-dominated systems \citep{k93,kk04}. Thus it makes sense that
they be flatter. The relative flatness of a bulge to its associated
outer disk has also been suggested as a pseudobulge indicator
(KK04). We will test this hypothesis in this paper by comparing the
flatness of bulges with disk like morphologies to that of bulges with
morphologies like those of elliptical galaxies.

The properties of pseudobulges are in stark contrast to the expected
end result of the hierarchical merging process; one does not expect
violent relaxation to produce spiral structure and dynamics that is
dominated by ordered motion. Further scenarions for the formation of
bulges have been suggested. Clump instabilities in disks at high
redshift can form bulge-like structures in simulations
\citep{noguchi99}. It is also plausible that gas rich accretion could
form dynamically cold bulges. Internal evolution of disks can drive
gas and stars to the center of a disk galaxy as well. The population
of bulges as a whole and any one particular bulge may be the result of
more than one of these processes.

However, the connections between bulge and disk stellar populations
\citep{peletier1996,macarthur2004}, inter-stellar medium
\citep{regan2001bima,fisher2006} and scale lengths
\citep{courteau1996} may suggest that pseudobulges form through
processes intimately linked to their host disks.  Furthermore,
\cite{droryfisher2007} find that classical bulges occur in
red-sequence galaxies and pseudobulges occur in blue cloud
galaxies. \cite{kk04} reviews the case that pseudobulges are not the
result of major mergers, but rather that internal disk-evolution may
be responsible for them. However, that bulges can form out of the
internal evolution of a disk is not a new idea; see, for example,
\citet{hohl1975}.

Simulations suggest that if disk galaxies do not experience major
mergers, they may evolve by redistribution of energy and angular
momentum driven by non-axisymmetries such as bars, ovals, and spiral
structure \citep{simkin1980,pfenniger1991,debattista2004}, resulting
in star formation and bulge-like stellar densities, thus forming
pseudobulges. Indeed, a correlation between central star formation
rate and the presence of bars and ovals has been detected
\citep[e.g.][]{sheth2005,jogee2005,fisher2006}. Also,
\cite{peeples2006} find that galaxies with nuclear rings and/or
nuclear spirals are more strongly barred. Thus we also investigate the
possible connection between driving mechanisms and structural
properties.
\begin{figure*}[t]
\begin{center}
\includegraphics[width=0.9\textwidth]{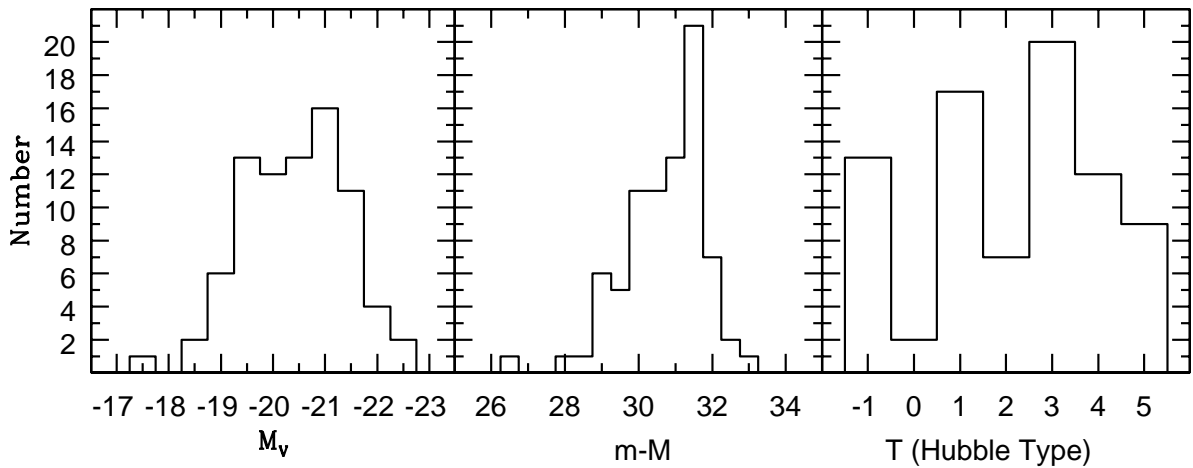}
\end{center}
\caption{The distribution of absolute magnitude (left), distance
  modulus (middle), and Hubble Type (right) for all 77 galaxies in our
  sample. \label{fig:sample}}
\end{figure*}

It is, thus, possible that the absence of a classical bulge in a
galaxy indicates that the galaxy has not experienced a major merger
since the formation of the disk. In this context, pseudobulges may be
thought of as more similar to pure disk galaxies that have a surface
brightness profile which breaks from the outer exponential profile to
a more steep inner surface brightness profile.  Though the frequency of
pseudobulges has not yet been robustly calculated, if they are common
then this implies that many disk galaxies did not suffer major mergers
since their formation.

This paper is organized as follows. In \S~2 we present the
observational data we use, and we discuss the surface brightness
fitting procedure. In \S~3 we present results on the location of
pseudobulges and classical bulges in various structural parameter
correlations. In \S~4 we discuss the flatness of pseudobulges and
classical bulges. In \S~5, we discuss behavior of different bulge
morphologies (nuclear bars, nuclear spirals, and nuclear rings) in
various parameter correlations. In \S~6 we summarize and discuss these
results. Finally, the appendix includes an image of each galaxy, all
decompositions and a discussion of the robustness of our decomposition
and fitting procedure.

\section{Methods and Observations}

\subsection{The Sample}

The aim of this work is to establish whether or not pseudobulges --
recognized by the presence of disk-like morphological features as
motivated and discussed in KK04 -- can be distinguished from classical
bulges simply by structural features in their surface brightness
profiles, most prominently their profile shape. Thus we ask whether
bulges that contain disk-like morphologies (pseudobulges) have lower
S\'ersic index and higher flattening ratios than bulges with
elliptical-like morphologies (classical bulges). Answering this
question requires high resolution imaging (preferably in the optical
bands) to detect the nuclear spirals, bars, and rings; and we need
surface brightness profiles with large dynamic range in radius to
accurately determine the parameters in Eq.~1 for a bulge-disk
decomposition.

We choose galaxies observable from the northern hemisphere that have
data in the {\em Hubble Space Telescope} (HST) archive. We limit our
selection to galaxies closer than $\sim 40$~Mpc to resolve features in
the bulge.

The link between non-axisymmetries (barred and oval distortions) and
secular evolution motivates us to create a sample containing roughly
equal numbers of galaxies with a driving agent (galaxies with a bar
and/or an oval) and galaxies without a driving agent. Detection of
oval distortions are discussed in \cite{k82}. They are identified by
nested shelves in the surface brightness profile usually having
different position angles. We identify bars by consulting the Carnegie
Atlas of Galaxies \citep{carnegieatlas} and the RC3 \citep{rc3}. If a
galaxy has both a bar and an oval, we call that galaxy
barred. Additionally, we look for bars and ovals in all galaxies using
K-band images from 2MASS \citep{2mass}.  Note that we do not
distinguish grand design spirals as a possible secular driver, though
they may be able to generate a similar but less extreme effect as bars
do (KK04). We use 39 undriven (no bar and no oval) galaxies and 38
driven galaxies (30 barred and 8 ovaled), a total of 77 galaxies.

In Fig.~\ref{fig:sample} we show the distribution of global properties
of the galaxies in our sample; these are also listed in Table 1. The
distribution of the distances of the galaxies in our sample is heavily
peaked at 16 Mpc due to the Virgo cluster and has a standard deviation
of 6 Mpc. We derive total magnitudes by 2D integration of our surface
brightness profiles. The distribution of absolute V magnitudes ranges
mostly from -19 to -22 with a median value of -20.5.

KK04 compiles data from several different studies to generate
preliminary statistics on how the frequency of pseudobulges varies
along the Hubble Sequence. They suggest that both pseudobulges and
classical bulges exist at intermediate Hubble Types (S0 to Sbc).
There appears to be a transition from classical bulges being more
frequent at early types (S0-Sb) to pseudobulges being more frequent at
later types (Sbc - Sd). They further suggest that classical bulges
will be almost non-existent at Hubble types Sc and later. In the right
panel of Fig.~\ref{fig:sample}, we show the distribution of Hubble
Types (taken from \citealp{carnegieatlas}) of the galaxies in our
sample. To test for differences between pseudobulges and classical
bulges we choose to sample the range of Hubble type from S0 to Sc. If
we combine this with our choice of evenly sampling driven and undriven
galaxies, we expect that our sample should overemphasize pseudobulges.
The distribution of Hubble types is as follows: 13 S0, 2 S0a, 18 Sa, 6
Sab, 20 Sb, 11 Sbc, 9 Sc. We will also use 24 E type galaxies from
\cite{kormendy2006virgo} as early-type sample in some parameter
correlations.
\begin{figure*}
\begin{center}
\includegraphics[height=0.9\textheight]{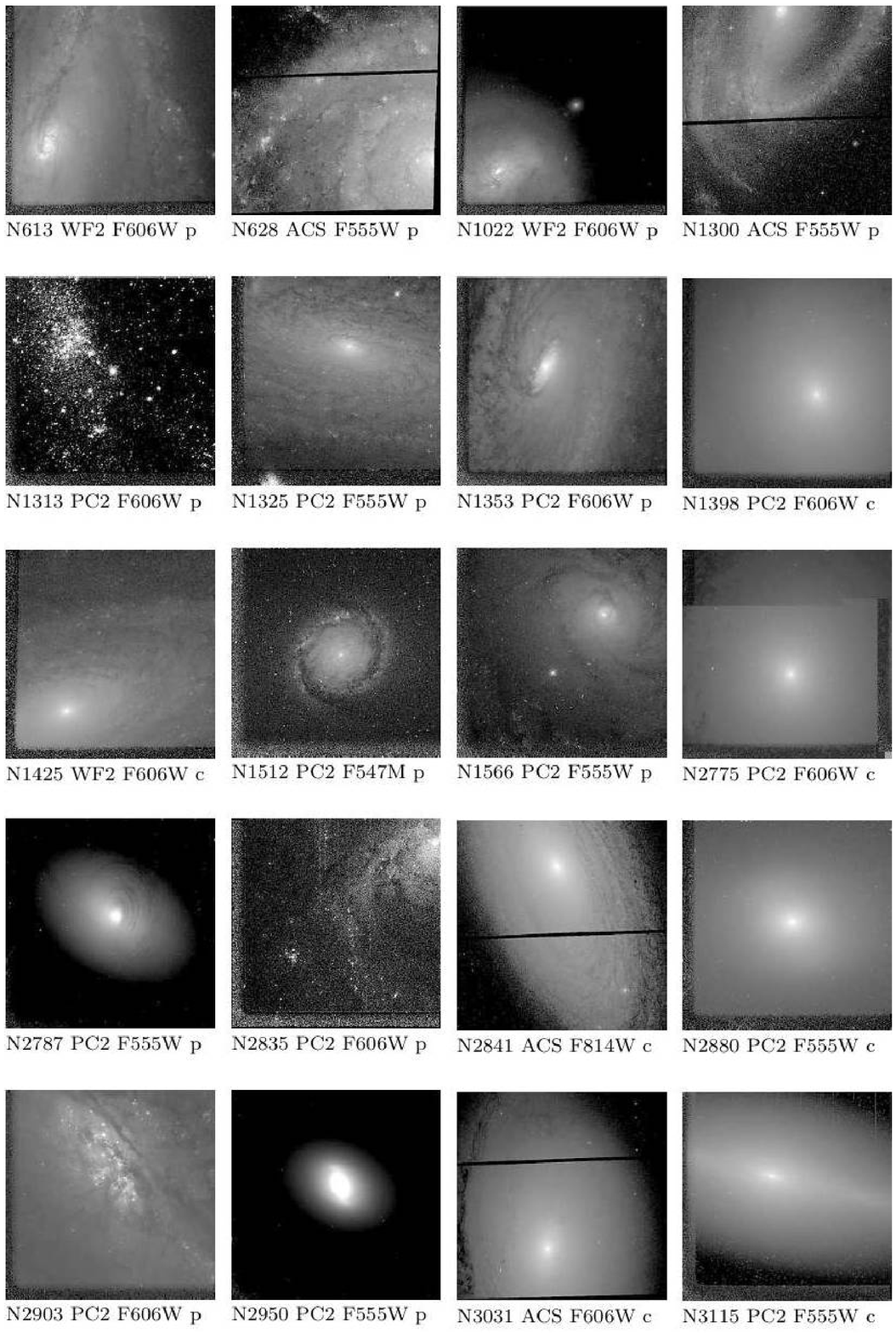}
\end{center}
\caption{High-resolution {\em HST} images of all bulges in our sample.
  The images are scaled logarithmically, and the range of shown
  intensities is chosen to emphasize those features which motivate the
  bulge classification. For all galaxies we give the galaxy name,
  imaging instrument used, filter, and the classification. We label bulges classified
  as classical by 'c' and pseudobulges by 'p'.
  The images show the inner 40 arcsec (ACS), 36 arcsec (PC2),
  and 80 arcsec (WF2) of the galaxies.
  \label{fig:bulgeid} }
\end{figure*}

\setcounter{figure}{1}
\begin{figure*}
\begin{center}
\includegraphics[height=0.9\textheight]{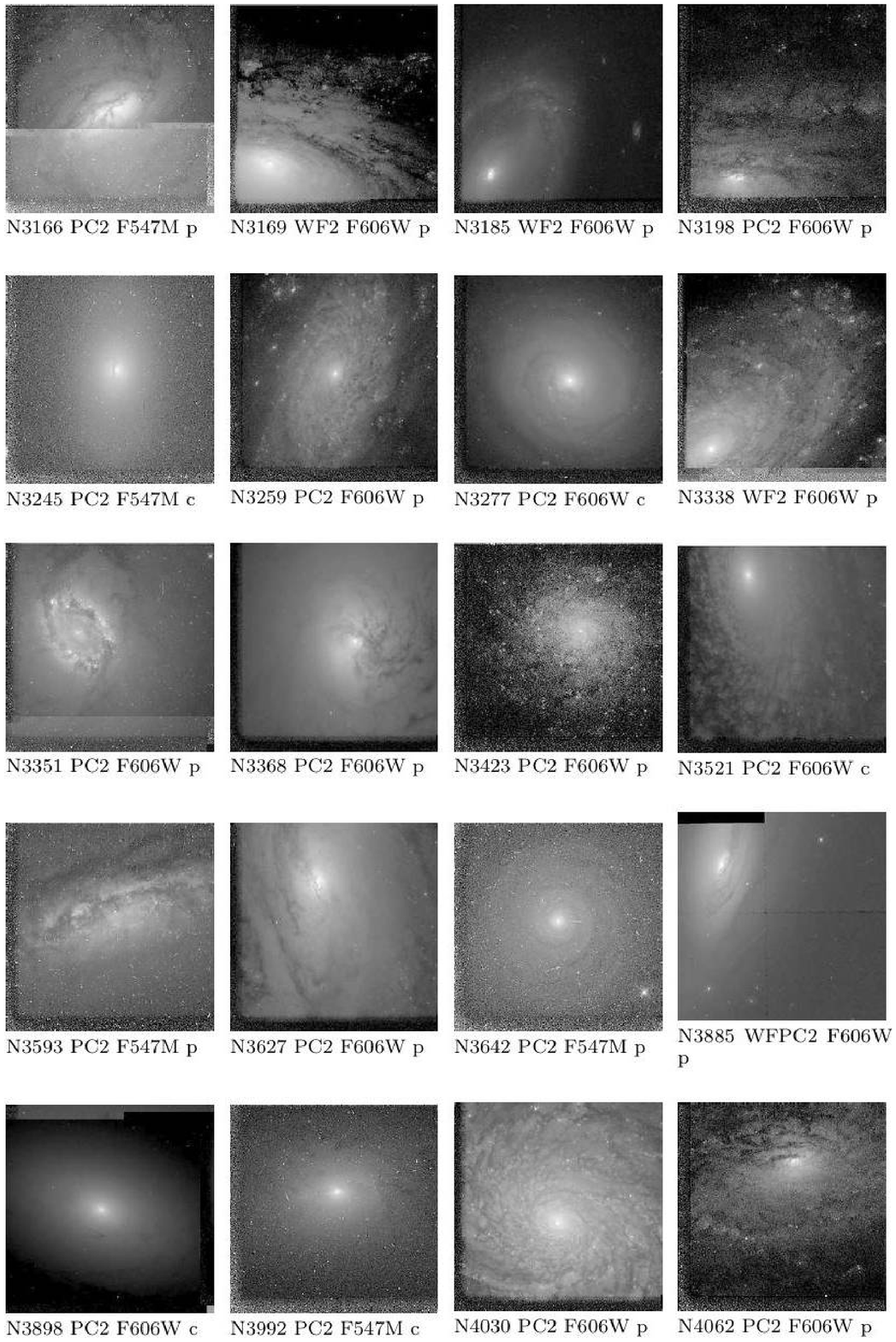}
\end{center}
\caption{Cont.}
\end{figure*}

\setcounter{figure}{1}
\begin{figure*}
\begin{center}
\includegraphics[height=0.9\textheight]{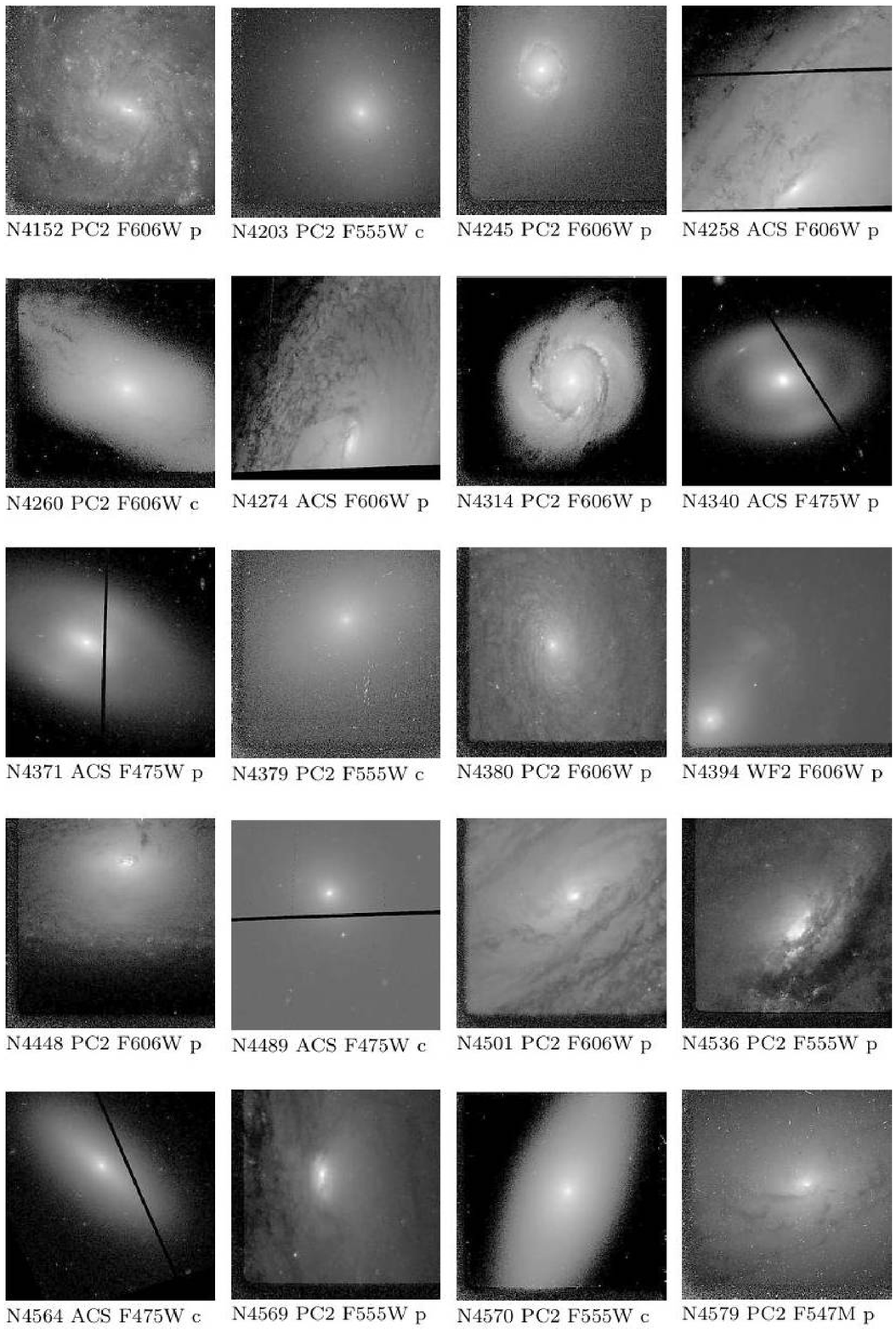}
\end{center}
\caption{Cont.}
\end{figure*}

\setcounter{figure}{1}
\begin{figure*}
\begin{center}
\includegraphics[height=0.9\textheight]{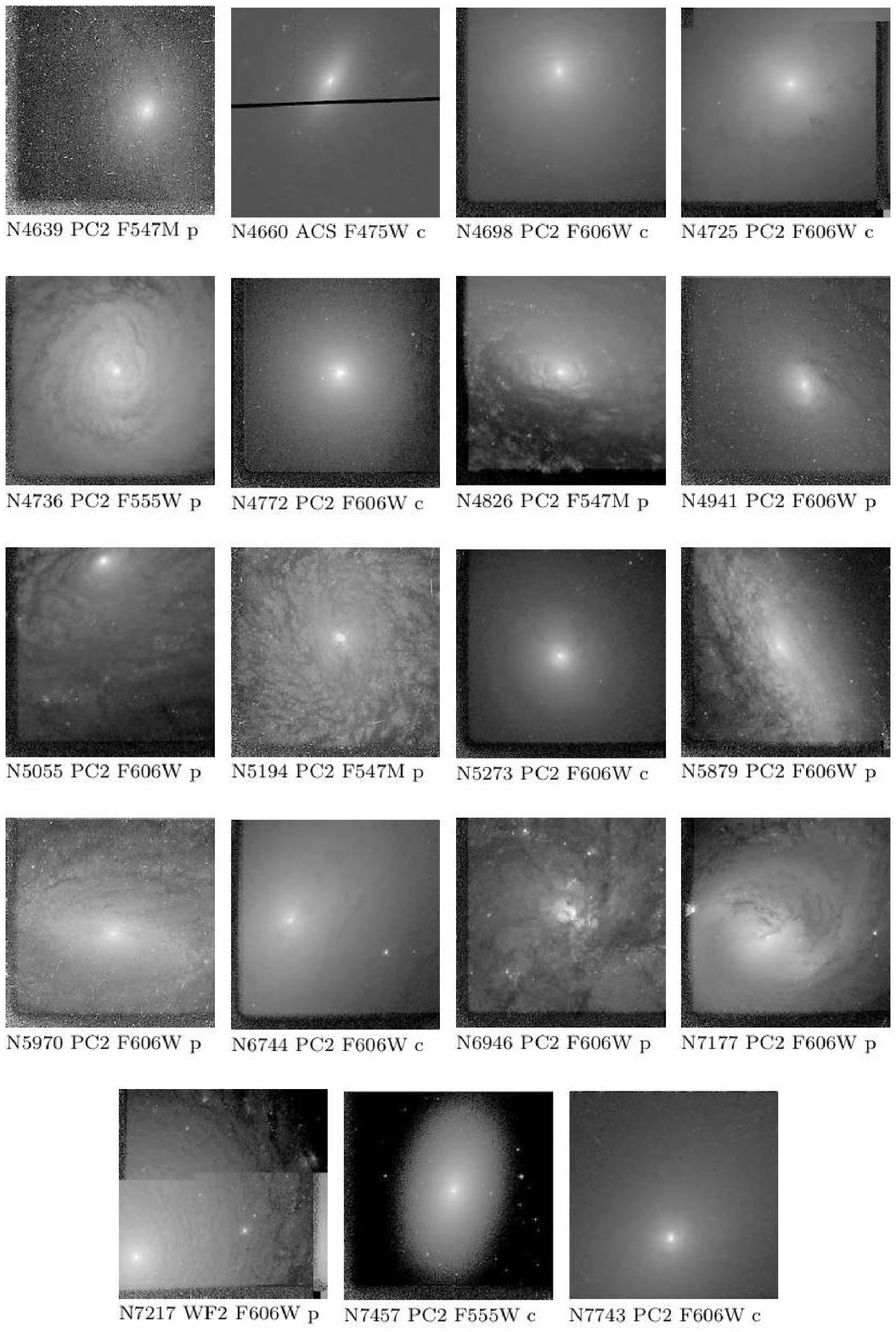}
\end{center}
\caption{Cont.}
\end{figure*}

\subsection{Identification of pseudobulges}

We define ``bulges'' photometrically, as excess light over the inward
extrapolation of the surface brightness profile of the outer disk. The
region of the galaxy where this excess light dominates the profile is
the bulge region. We classify galaxies as having a pseudobulge by
their morphology within this bulge region; if the bulge is or contains
any of the following features: a nuclear bar, a nuclear spiral, and/or
a nuclear ring, then the bulge is called a pseudobulge.  Conversely,
if the bulge better resembles an elliptical galaxy (relatively
featureless isophotes), then the bulge is called a classical
bulge. This method is discussed in KK04. The existence/absence of
visibly identifiable disk-like structure in a bulge correlates with
properties of the bulge and the whole galaxy. The same method is shown
to be successful in identifying bulges with higher specific star
formation rates \citep{fisher2006} and globally bluer galaxies
\citep{droryfisher2007}.

\begin{figure*}
\begin{center}
\includegraphics[width=0.48\textwidth]{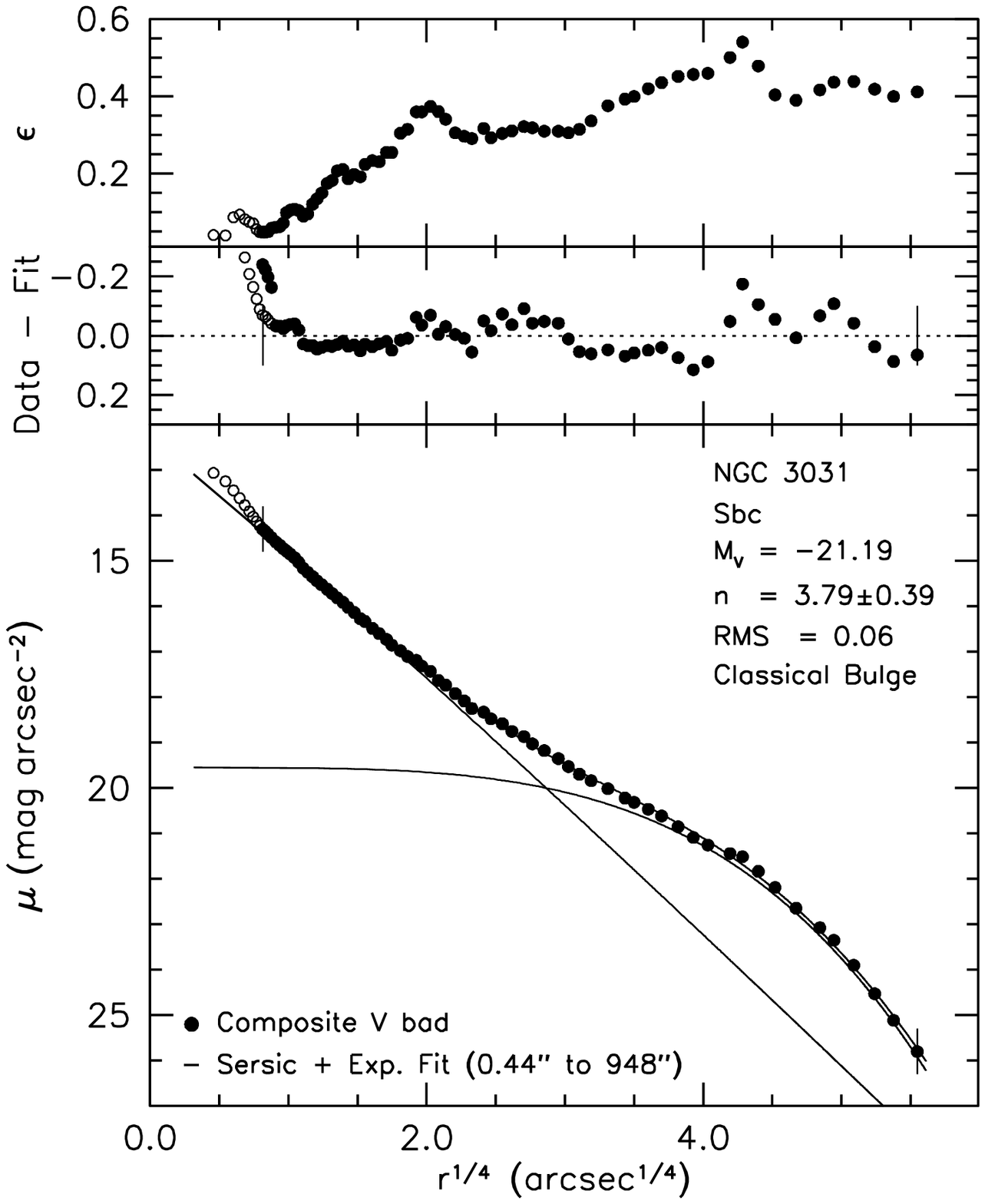}
\includegraphics[width=0.48\textwidth]{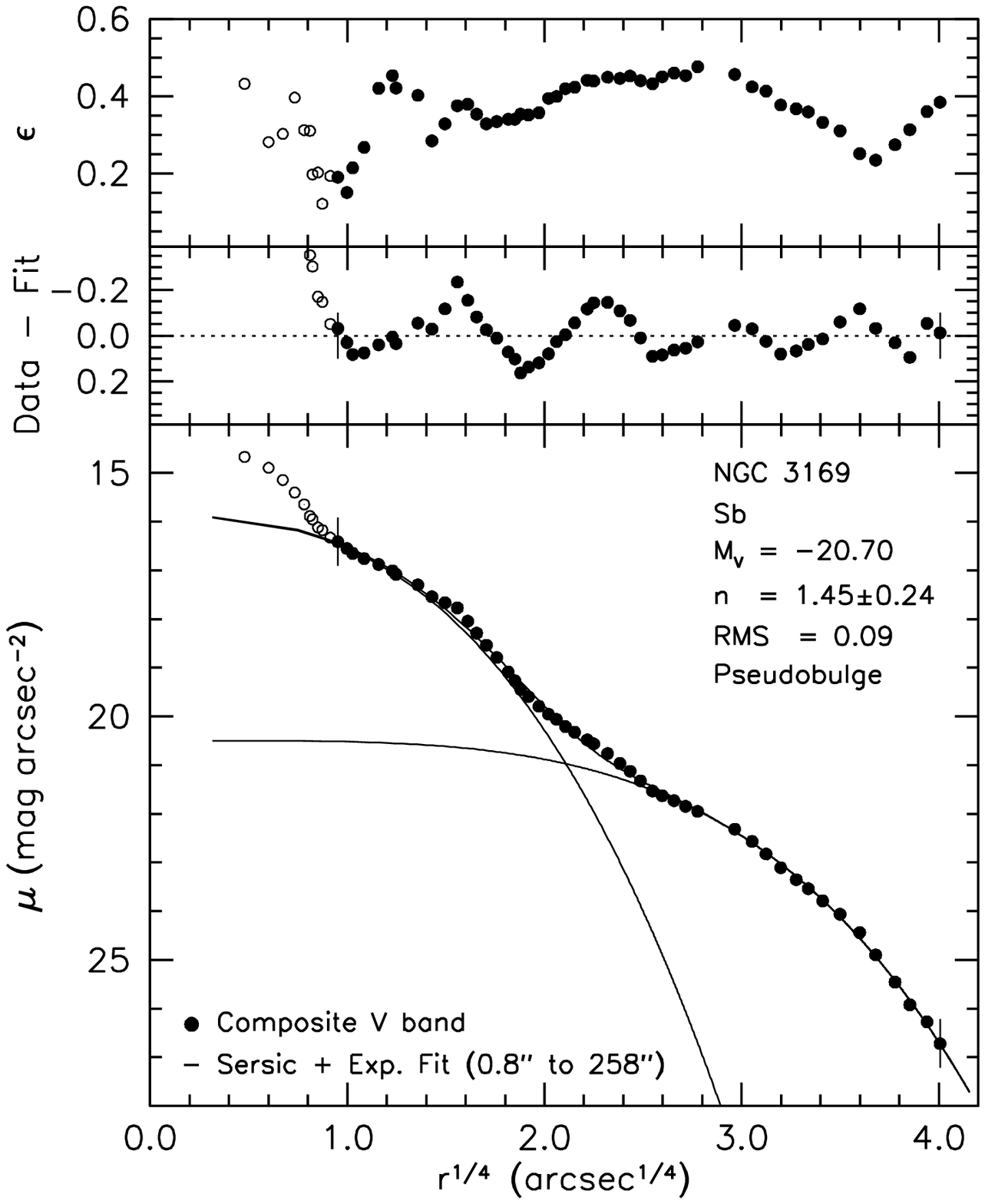}
\includegraphics[width=0.48\textwidth]{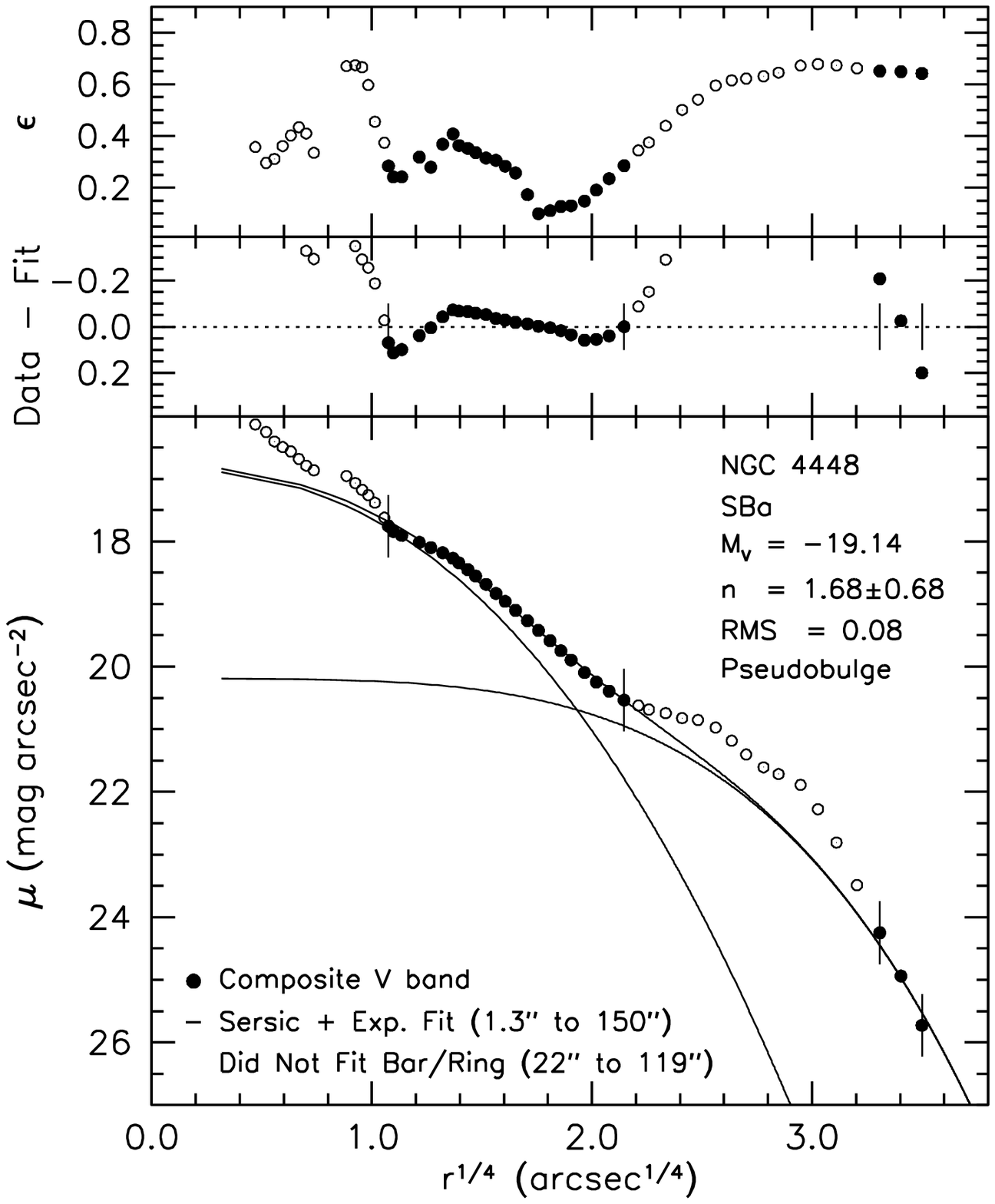}
\includegraphics[width=0.48\textwidth]{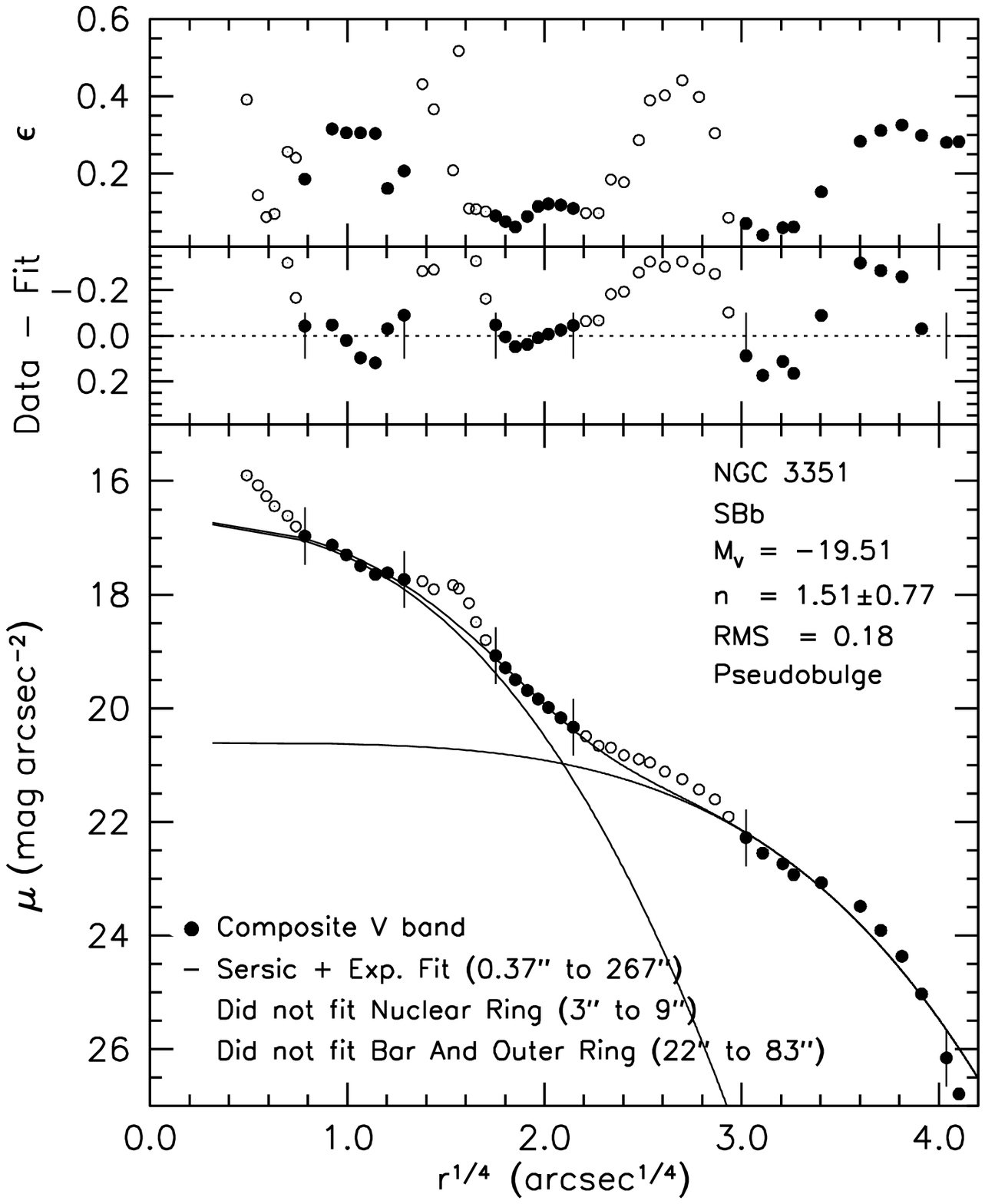}
\end{center}
\caption{We show four example decompositions, spanning the range of
  the types of profiles that we find in the galaxies investigated in
  this paper. All data are plotted against $r^{1/4}$ as this
  projection emphasizes deviations from the fit better than $\log(r)$,
  and shows the central regions better than plotting against linear
  radius would.  For each galaxy we show a profile (black dots) of
  (from bottom to top panel) V band surface brightness; residual
  profile of the data minus $-2.5\log I(r)$, where $I(r)$ is the fit
  of the data to Eq.~1; and ellipticity profile, where
  $\epsilon=1-b/a$, and $b/a$ is the axial ratio of the galaxy. In the
  bottom panel we also show the decomposition: the inner S\'ersic
  profile and the outer exponential profile as black lines. The
  vertical hatches on the profile indicate the boundaries of the
  regions included in the fit, the bottom caption also notes the total
  range of the fit. Excluded data regions are marked by open symbols,
  fitted regions by closed symbols. The caption gives the galaxy name,
  the Hubble type, the total absolute magnitude, the S\'ersic index of
  the bulge and the root-mean-squared deviation of the fit to the fit
  region of the data. \label{fig:egprof}}
\end{figure*}

Fig.~\ref{fig:bulgeid} shows high-resolution {\em HST} images of the
bulge region of all galaxies in our sample. All images are taken in
close-to $V$ band filters.  Note first that classical bulges (for
example NGC~1398, NGC~2775, NGC~2880, and NGC~3115, all in the
rightmost column of the first page of the figure) have a smooth
stellar light profile. There is no reason evident in the images to
think that any of these galaxies harbor a pseudobulge. The bulges fit
the description of E-type galaxies.  On the other hand, NGC~4030 shows
a face-on nuclear spiral (bottom row on the second page). The spiral
dominates the radial profile for more than a kiloparsec. NGC~4736
(fourth page, second row) also has a nuclear spiral pattern but also
has a nuclear bar; note how the spiral arms seem pinched vertically in
the image. \cite{mollenhoff1995-n4736} study this nuclear bar in more
detail using dynamical modeling. NGC~4371 (third page, middle row) is
an S0 galaxy with at least one nuclear bar. NGC~3351 (second page,
third row) has a prominent nuclear ring that heavily distorts the
surface brightness profile (see Fig.~\ref{fig:egprof}); this nuclear
ring is quite well known \citep{hubbleatlas}. The bulge of NGC~2903
(first page, bottom row) shows a chaotic nuclear region; it appears
nearly spiral but is not regular enough over a significant radial
range to call it a nuclear spiral.

In this paper we classify bulges with near-$V$-band images (F547M,
F555W, \& F606W). Thus our method is subject to the effects of dust
obscuration. However, the structures used to identify pseudobulges are
usually experiencing enhanced star formation rates \citep{fisher2006}.
Pseudobulges are, therefore, easier to detect in the optical band
passes where the mass-to-light ratios are more affected by young
stellar populations, rather than in the near infrared where the
effects of dust are less pronounced. It is important to note that
classical bulges may have dust in their center, as do many elliptical
galaxies \citep{Laueretal05}. In fact, many classical bulges shown in
Fig.~\ref{fig:bulgeid} have some dust lanes in their bulges. The
presence of dust alone is not enough to classify a galaxy as
containing a pseudobulge; instead, it must be of a disk-like nature.

\subsection{Data Sources and Surface Photometry}

As stated in the introduction, the S\'ersic function provides better
fits and more information over two-parameter fitting functions
describing surface brightness profiles of bulges and elliptical
galaxies. However, this information comes at the price of more
detailed observations. \cite{saglia1997} shows that replacing an
$r^{1/4}$-model with a S\'ersic model makes little difference in
residuals, unless one has data with high dynamic range in
radius. Further, the coupling between parameters in the S\'ersic
function can be quite high \citep{graham1997}. Thus again, it is
necessary to fit the decomposition (Eq.~1) to large radial range in
order to minimize these degeneracies. For each galaxy we therefore
combine multiple data sources together: high-resolution HST imaging in
the center, and ground based wide-field images covering the outer
disk. Comments on data sources follow. Table~1 lists the sources of
data used for each galaxy.
\begin{figure}[t]
\begin{center}
\includegraphics[width=0.4\textwidth]{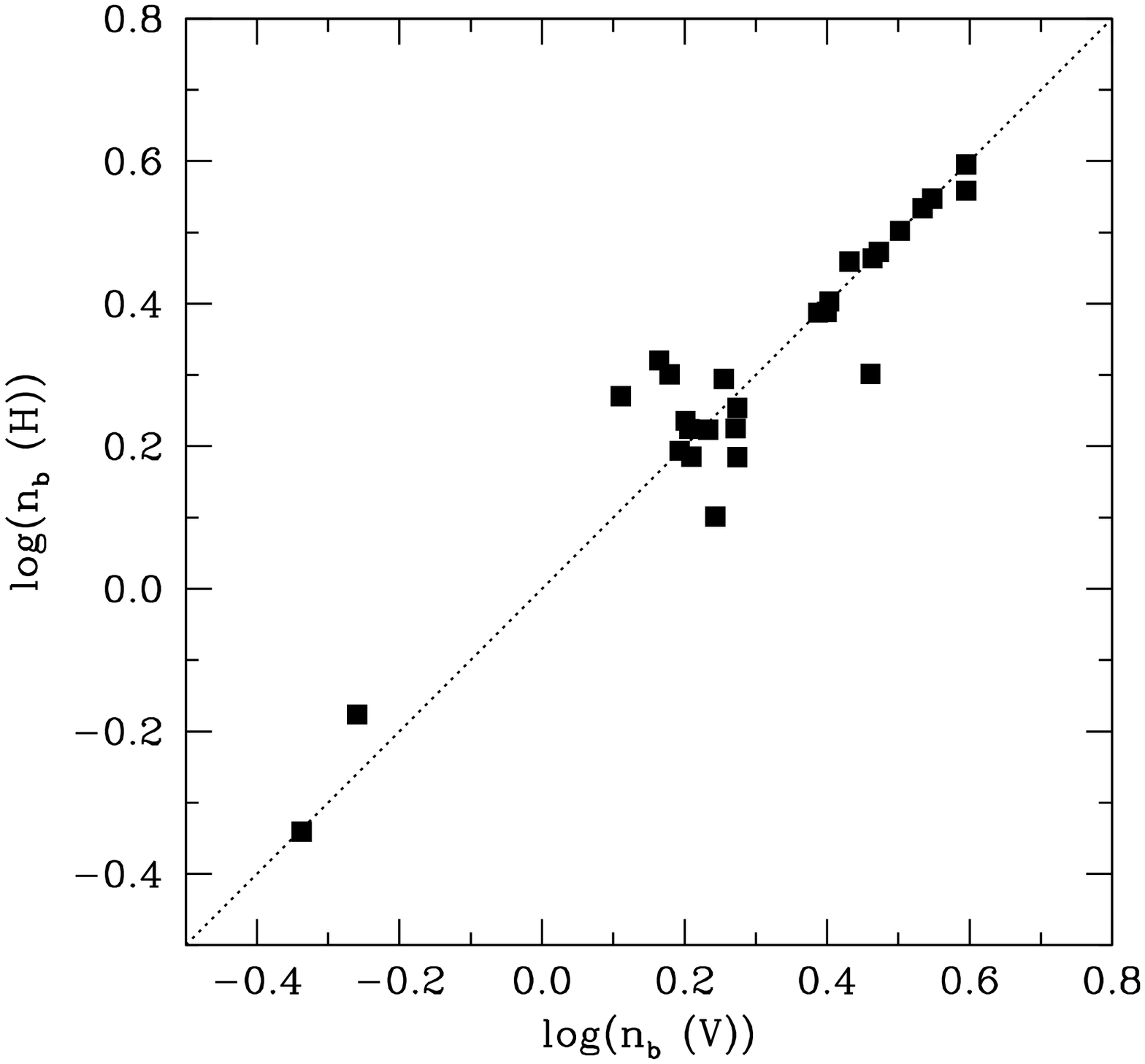}
\includegraphics[width=0.4\textwidth]{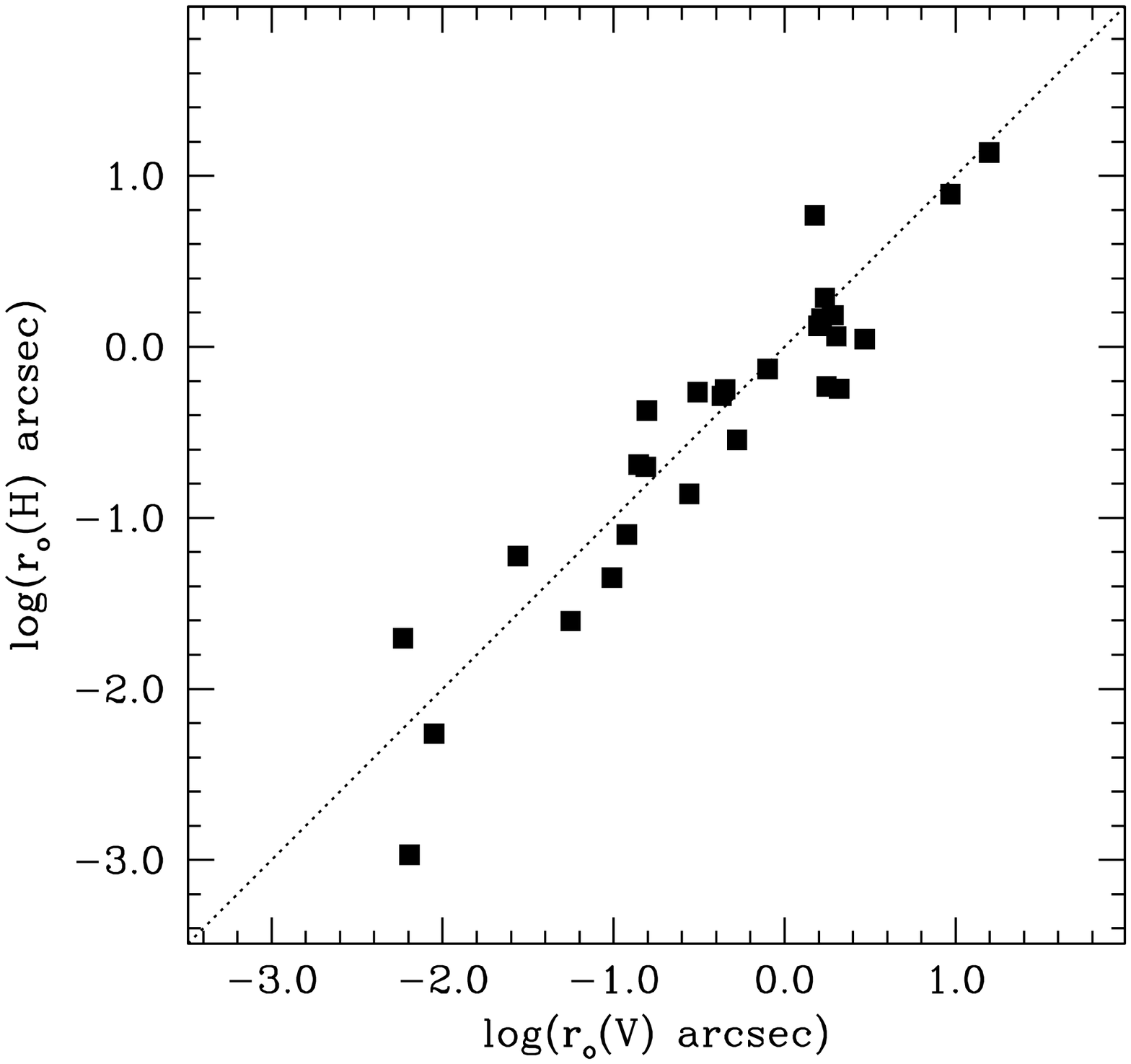}
\end{center}
\caption{Comparison of structural parameters ($r_0 \& n_b$) derived
  from bulge-disk decompositions in the $H$-band to those in the $V$-band. The
  identity relation is marked by the dashed line.\label{fig:nir}}
\end{figure}

All profiles contain HST data sources. PC2 data has a small
field-of-view ($\sim18\times 18$~arcsec$^2$); thus, it is critical to
supplement PC2 data with wide field data. ACS/WFC has proven to be an
excellent instrument for obtaining large radial fitting range. It
provides a reasonable sized field-of-view ($\sim100\times
100$~arcsec$^2$) at high spatial resolution (0.049 arcsec
pixel$^{-1}$).

For as many galaxies as possible we obtain wide field images from the
Prime-Focus-Camera on the McDonald 0.8~m telescope. This instrument
provides a large unvignetted field of view ($45\times 45$ arcmin$^2$),
and a single CCD detector. Therefore, we can more accurately carry out
sky-subtraction.  These data generally are the deepest. We also
include images from the Sloan Digital Sky Survey \citep{sdssdr4},
2Micron All Sky Survey (2MASS) and the Isaac Newton Group (ING)
Archive. Individual data sources are noted in Table~1.

All raw data (McDonald 0.8~m \& ING data) are bias-subtracted,
flat-fielded, and illumination corrected. We subtract the sky
background by fitting a plane to a smoothed version of the image where
the galaxy and bright stars have been removed.

We calculate Johnson $V$-band magnitude zero points using the
transformations in \cite{holtzman1995} for the WFPC2 images and
\cite{sirianni2005} for the ACS images. SDSS g and r profiles are
converted to a single V-band profile for each galaxy using the
transformations in \citet{smithetal2002}. Other profiles are simply
shifted to match in surface brightness.  These transformations are
derived on galactic disk stars, not galaxies.  Further, the
calculations require color information. We use colors from Hyper-LEDA,
which refer to colors of the entire galaxies, and the galaxies in our
sample most certainly have non-zero color gradients.  Therefore the
absolute values of surface brightness in this paper are not expected
to be consistent to more than 0.3~mag. However, this does not affect
our conclusions which are based the structure in the profiles and not
on absolute magnitude. We check that our total magnitudes are
consistent with those published in the RC3 and Hyper-LEDA.

We use the isophote fitting routine of \cite{bender1987}. We identify
and mask interfering foreground objects in each image. Then we fit
ellipses to isophotes by least squares.  Here, isophotes are sampled
by 256 points equally spaced in an angle $\theta$ relating to polar
angle by $\tan \theta = (a/b)\,\tan \phi$, where $\phi$ is the polar
angle and $b/a$ is the axial ratio. The software determines six
parameters for each ellipse: relative surface brightness, center
position, major and minor axis lengths, and position angle along the
major axis.

\subsection{Bulge-Disk Decomposition}

We carry out a bulge-disk decomposition on each galaxy in our sample
by fitting Eq.~1 to the major axis surface brightness profiles. Our
decomposition code is also used by \cite{kormendy2006virgo} to fit S0
and E-type galaxies with the disk component ``turned off''. The
average root-mean-square deviation is $\sim0.09\pm0.03$ mag
arcsec$^{-2}$. The largest deviation of any of the data from its
fitted profile is $0.18$ mag arcsec$^{-2}$.

The parameters determined in bulge-disk decomposition with a S\'ersic
model bulge are coupled. \cite{macarthur2003} shows that if the bulge
is sufficiently large and the resolution is sufficiently small,
initial parameter estimates do not affect the final fit too much. For
the most part, this is true. Yet, our experience is that initial
parameter estimates may still affect the resulting fit. For each
galaxy, we begin with a large parameter range that is symmetric about
the initial guess. Then we re-fit the galaxy iteratively adjusting the
range of allowed parameters to be narrower. This results in slight
changes of best-fit values with lower $\chi^2$ than without this
iteration. The details of each profile are used to decide the width of
available parameter space each parameter is given. Typically the
available range for $n_b$ is $\Delta n_b \sim 2-3$.

The decomposition is carried out on a major axis profile using the
mean isophote brightness. It does not take ellipticity into account
during the fitting. However, these galaxies are known to have varying
ellipticity profiles \citep{k93,fathi2003}. Thus, we take the mean
ellipticity for each component and adjust the luminosity accordingly:
$L=(1-\bar{\epsilon})L_{\mathrm{fit}}$. The radius of the component is defined
as the radius range within which that component dominates the light of
the profile. We also adjust all magnitude dependent quantities
(luminosity and surface brightness) for Galactic reddening according
to \cite{schlegel}.  We use the distances of \cite{tully1998}. We do
not make any corrections for extinction within the galaxies being
studied.

We carry out the bulge-disk decompositions in the $V$ band. This
ensures that our results will remain applicable to large surveys
commonly done in the optical bands (e.g.\ SDSS). However, the radial
variation in age and metallicity that exists within galaxies may bias
the parameters of bulge-disk decomposition. \cite{carollo2002} shows
that the centers of intermediate type galaxies contain significant
structure in $V-I$ color maps, and this variation occurs on scales
smaller than the bulge. Thus, there is doubt as to whether the
parameters derived on profiles in the middle optical band passes truly
reflect the properties of the stellar mass distribution.

It is beyond the scope of this paper to investigate the correlations
of colors with S\'ersic parameters. Yet, as a check of the stability
of structural quantities compared to color gradients, we carry out
bulge-disk decompositions on all galaxies in our sample that have
archival NICMOS images in the filter $F160W$ ($H$ band). We supplement
those data with ground-based wide-field data. For this purpose we use
mainly 2MASS data and any H-band data available in NED. We then
compare the fit parameters in $V$ to those in $H$. The results are
shown in Fig.~\ref{fig:nir}. In those parameters which do not depend
on magnitude ($r_0$ and $n_b$) there is little difference. The average
difference is $\Delta_{_{V-H}}n_b\sim-0.03$, and
$\Delta_{_{V-H}}\log(r_0)\sim 0.07$. This is similar to the results of
other papers that have done decompositions at multiple wavelengths
(e.g.~\citealp{graham2001,macarthur2003}).

In Fig.~\ref{fig:egprof} we show four examples of our photometry and
the bulge-disk decompositions. For each galaxy we show an ellipticity
profile, a residual profile, and a surface brightness profile along
with the decomposition determined by fitting Eq.~1 to the galaxy's
surface brightness profile. These examples are selected to show the
range in typical profiles in this paper. The entire sample of fits is
shown in the appendix.

The top two panels of Fig.~\ref{fig:egprof} shows two galaxies
(NGC~3031, left panel; NGC~3169, right panel) that are well described
by an inner S\'ersic function and an outer exponential disk. In both
profiles the fit covers a dynamic range of $\sim$12 mag
arcsec$^{-2}$. NGC~3031 is well fit (RMS=0.06 mag arcsec$^{-2}$) by
Eq.~1 over $\Delta \log(r/\mathrm{arcsec})=3.3$ and NGC~3169 for
$\Delta \log(r/\mathrm{arcsec})=2.51$. Deviations are small, and there
is little-to-no substructure evident in these profiles. Note that each
bulge has a nuclear excess of light over the inward extrapolation of
the fit, and that this nucleus is excluded from the fit (vertical
hatches indicate beginning and end of fit range and excluded data
points are marked by open symbols).

Despite its successes, the S\'ersic bulge plus outer exponential disk
model of bulge-disk galaxies does not account for many features of
galaxy surface brightness profiles. Disks of intermediate type
galaxies commonly have features such as bars, rings, and lenses (see
\citealp{k82} for a description of these). Further, \cite{carollo2002}
shows that many bulges of early and intermediate type galaxies contain
nuclei. The bottom panels of Fig.~\ref{fig:egprof} shows two such
examples of galaxy profiles with significant deviations from the
fitted decompositions. NGC~4448 (bottom left) is an example of a
barred galaxy in which the bar is an especially prominent perturbation
to the outer exponential surface brightness profile. NGC~3351 (bottom
right) is an example of a complicated surface brightness profile, with
multiple substructures that are not well described by the smooth
nature of the bulge-disk model used here. This galaxy contains a
nuclear ring near $\sim 4$~arcsec, and a bar from about 20 to
80~arcsec. These galaxies are not well described by Eq.~1, yet we do
our best to decompose as many galaxies as possible.

Bars, rings, lenses, and similar features do not conform to the smooth
nature of Eq.~1, hence we carefully exclude regions of the profile
perturbed by such structures from the fit. This is a risky procedure,
as it requires selectively removing data from a galaxy's profile, and
undoubtedly has an effect on the resulting parameters. We are often
helped to identify bars by the structure of the ellipticity
profile. Notice the the peak in ellipticity near 50~arcsec in
NGC~3351. We try to err on the side of removing the fewest points
possible.  For those galaxies in which a bar is present, it is our
assumption that removing the bar from the fit provides the best
estimation of the properties of the underlying bulge and disk. We test
our method by removing a typical number of points from a few galaxies
with smooth profiles (NGC~2841, NGC~1425, NGC~4203, \& NGC~1325). The
result is that S\'ersic index can vary as much as $\Delta n_b\sim
0.5$, and characteristic radius by $\Delta \log(r_0)\sim 0.5$. The
variance of these two parameters is tightly coupled and this is
reflected by the uncertainty. Central surface brightness was mostly
unaffected. If a region is not included in a fit we show that in the
figure by using open symbols. A detailed discussion of the effects of
bars and other features on the surface brightness profiles is given in
the appendix.
\begin{figure}[t]
\begin{center}
\includegraphics[width=0.48\textwidth]{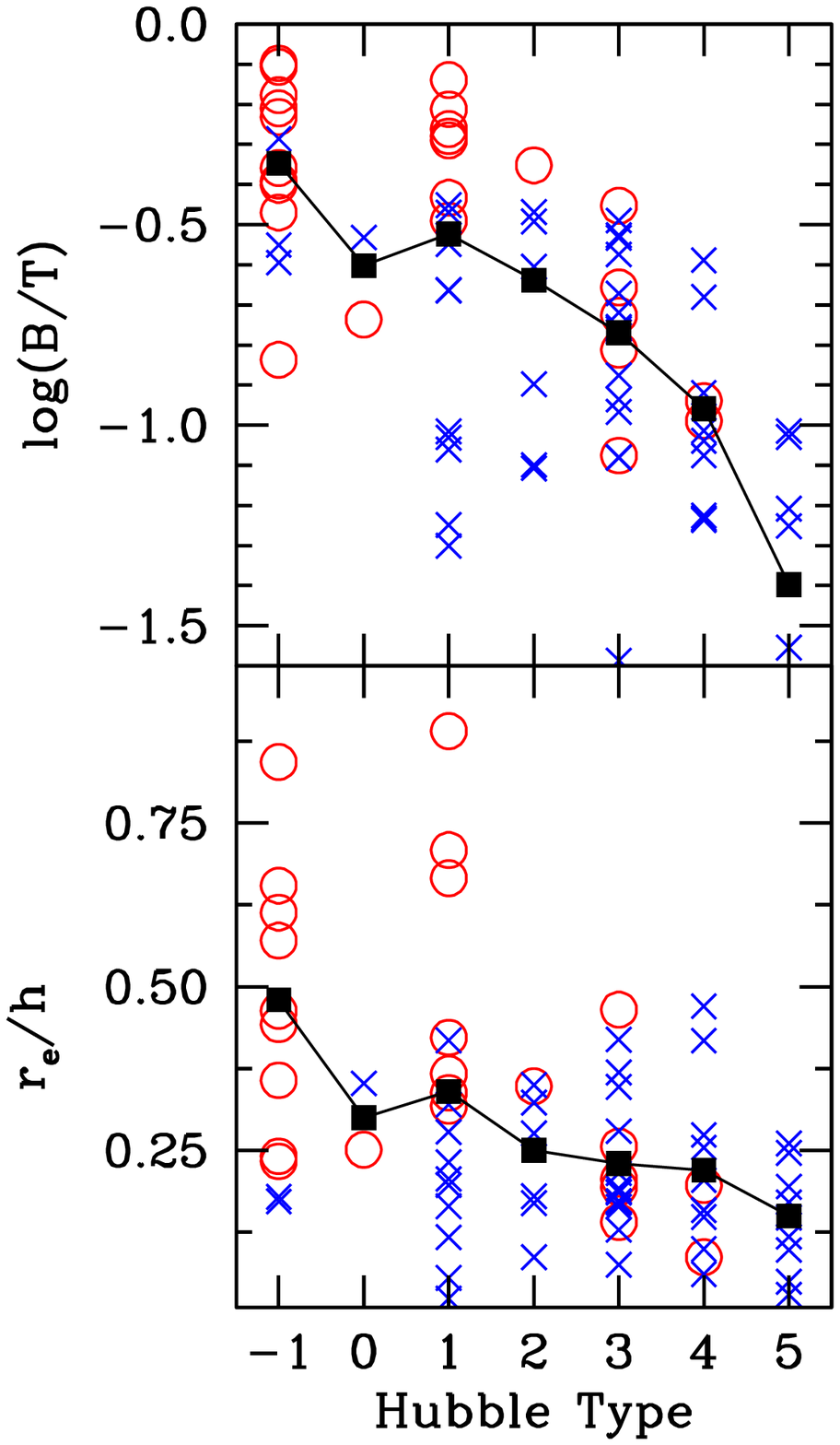}
\end{center}
\caption{The bulge-to-total ratio, $B/T$, top, and the ratio of the
  effective radius to the disk scale length, $r_e/h$, bottom, of
  classical bulges (red circles) and pseudobulges (blue x's) plotted
  against Hubble types. The black squares, connected by the black
  line, represents the average for all bulges (both pseudo- and
  classical) in each type. \label{fig:hubbletype}}
\end{figure}

\section{Structural Properties Of Pseudobulges And Classical Bulges}
\subsection{Bulge Prominence Of Classical \& Pseudobulges Along The
  Hubble Sequence}

The primary distinction between most previous studies and this one is
that we do not treat bulges as an homogeneous set of objects. Here, we
report on how morphologically identified pseudobulges and classical
bulges, taken to be distinct entities, behave in the parameters
obtained from bulge-disk decomposition. Further we wish to know if the
S\'ersic index is able to distinguish these two separate bulge types
as has been suggested by many authors and not yet systematically tested.

Fig~\ref{fig:hubbletype} shows how the bulge-to-total ratio
(luminosity of the bulge, divided by total luminosity of the fit;
hereafter $B/T$), and the ratio of the bulge half-light radius ($r_e$)
to the scale-length of the outer disk ($h$) both correlate with Hubble
type. We calculate the effective radius as
\begin{equation}
r_e=(b_n)^nr_0,
\end{equation}
where $b_n$ is a proportionality constant whose expansion given in
\cite{macarthur2003}.  In all figures red circles represent classical
bulges and blue crosses represent pseudobulges. In
fig~\ref{fig:hubbletype}, the connected black squares show the average
for each Hubble type.  There are only two S0/a (T=0) galaxies; thus
the dip in the average of $B/T$ could merely be small number
statistics. Also, a few of the Sc (T=5) galaxies are not shown due to
their very small $B/T$.
\begin{figure*}[t]
\begin{center}
\includegraphics[width=0.8\textwidth]{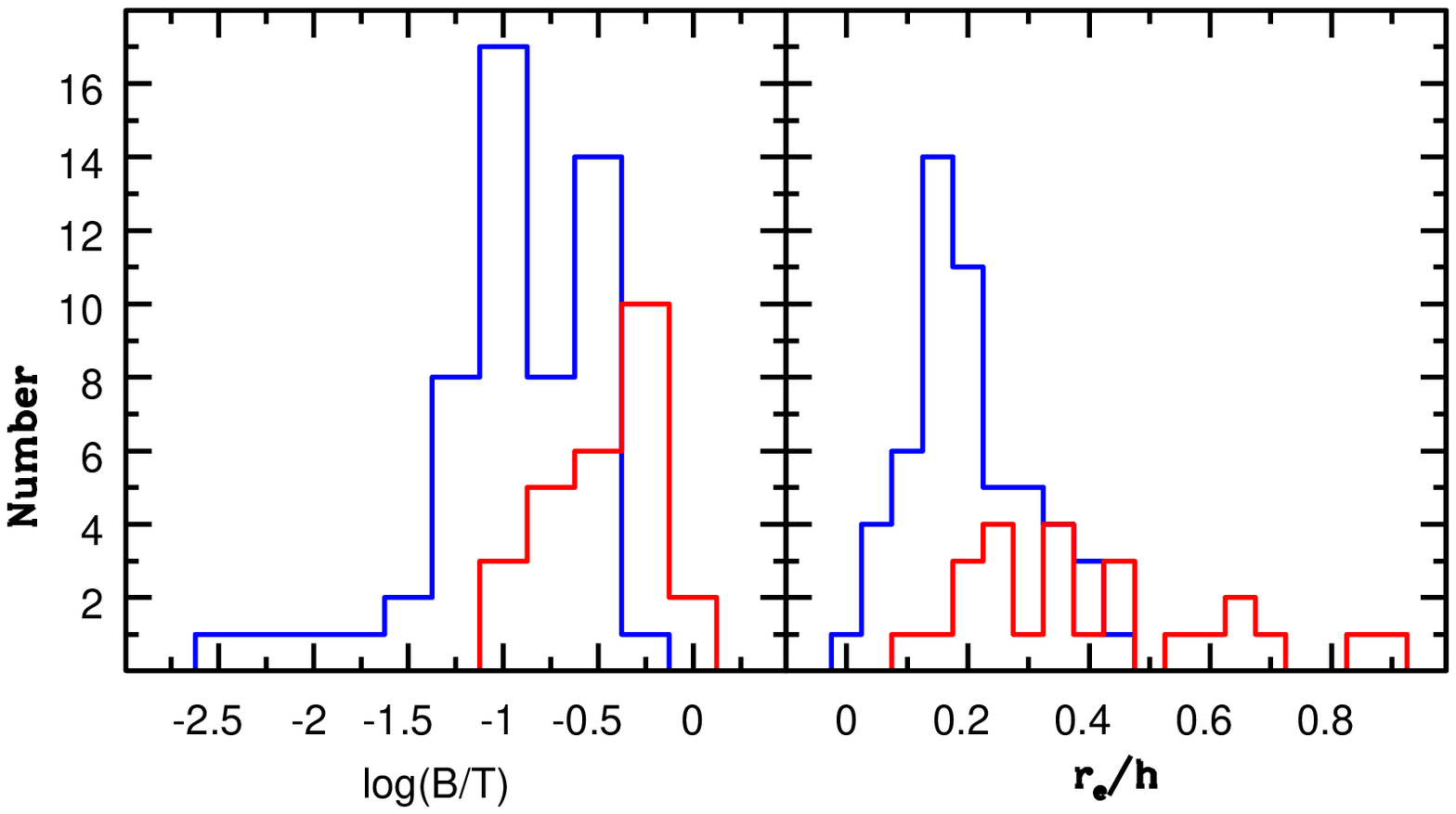}
\end{center}
\caption{Histogram of $\log(B/T)$ (left) and $r_e/h$ (right). Bin
  spacing is $\delta \log(B/T) = 0.25$ and $\delta (r_e/h)=0.05$. Blue
  lines represent pseudobulges and red lines are for classical
  bulges. \label{fig:bthist}}
\end{figure*}

As shown in the top panel of Fig.~\ref{fig:hubbletype}, bulge-to-total
ratio is a decreasing function of Hubble type. This is not surprising
as $B/T$ is part of the original classification criteria. This
behavior has been found by many authors, for a detailed study see
\cite{graham2001}. This at least confirms that our decomposition
method is sound, and able to recover the well established correlation.

Pseudobulges and classical bulges overlap in range of bulge
prominence, shown in the left panel of Fig.~\ref{fig:bthist}. The
average classical bulge has $\langle B/T \rangle = 0.41$. The width of
the distribution is 0.22. The average pseudobulge has $\langle B/T
\rangle = 0.16$. The width of the distribution is 0.11. Pseudobulges
are on average smaller fractions of the total luminosity of the galaxy
in which they reside, yet there is significant overlap (see also
\citealp{droryfisher2007}). This is no surprise, we would expect
pseudobulges to be smaller as they are thought to be products of
secular evolution of the disk; hence their luminosity is expected to
be limited to some fraction of the luminosity of the disk. On the
other hand, classical bulges form independently of their disk,
presumably through major mergers. The full range values of $B/T$ is
therefore possible.

NGC~2950 is a pseudobulge with $B/T=0.51$. It is hard to believe that
secular evolution can drive half of the stellar mass into the central
region of a galaxy. This galaxy is an SB0 with a prominent nuclear bar
and nuclear ring, also noticed by \cite{erwin2003}. Thus, by our
definition it is a good pseudobulge candidate. The next most prominent
pseudobulge has $B/T=0.35$; NGC~2950 is more than one standard
deviation away in the distribution of pseudobulge $B/T$ from the next
most prominent pseudobulge. We feel that NGC~2950 is an exceptional
case, and should not be taken as a normal pseudobulge. That it is an
S0 galaxy strengthens this interpretation. The unusually large $B/T$
may be a result of the processes which made the galaxy S0 (for example
gas stripping by ram pressure and/or harassment; see \citealp{MKLDO96}),
rather than secular evolution. Also, our analysis may be an over
simplification of the population of bulges, in that composite systems
(``bulges'' with both a pseudobulge and a classical bulge) may
exists. This could artificially increase the $B/T$ ratio.

\cite{courteau1996} show that the ratio of scale lengths for galaxies
when fitted with a double exponential (bulge and disk) is tightly
coupled. \cite{macarthur2003} find, with fits to S\'ersic plus outer
exponential profiles in galaxies spanning Hubble type Sab to Sd, that
the coupling is $\langle r_e/h \rangle = 0.22\pm0.09$, and that the
ratio becomes smaller toward later Hubble types (see also lower panel
of Fig.~\ref{fig:hubbletype}). It appears that the correlation of
$r_e/h$ with Hubble type may be driven primarily by the number of
classical bulges in each Hubble type; this statement is very uncertain
due to small numbers in each Hubble type, though.

The right panel of Fig.~\ref{fig:bthist} shows the distribution of
$r_e/h$ for the galaxies in our sample (red lines are for classical
bulges and blue lines represent the distribution of pseudobulges).  We
find that $\langle r_e/h \rangle = 0.28\pm0.28$ for all galaxies in
our sample, and that $r_e/h$ decreases toward later Hubble types. The
average is higher and the scatter is larger than in
\cite{macarthur2003}, most likely because our sample targets
earlier-type galaxies.  Considering bulge types separately, we find
that for pseudobulges $\langle r_e/h \rangle = 0.21\pm0.10$, and for
classical bulges $\langle r_e/h \rangle = 0.45\pm0.28$. The
distribution of $r_e/h$ for pseudobulges is clearly much narrower than
that of the classical bulges. Furthermore, $r_e$ and $h$ do not appear
to be correlated in classical bulges.

Our finding that the half-light radius of pseudobulges is well
correlated with the scale length of the outer disk, while the scale
length of classical bulges is not correlated with that of the disk is
consistent with the interpretation that $r_e \propto h$ is due to a
secular formation of pseudobulges. However, inspection of
Fig.~\ref{fig:bthist} shows that the range of $r_e/h$ over which we
find classical bulges is large and overlaps with the pseudobulges
significantly. Thus, $r_e/h$ may not be a good diagnostic tool for
identifying pseudobulges and classical bulges; however the finding
that the scale length of the disk is correlated with the size of its
pseudobulge but not correlated with classical bulges is a strong
indication that there indeed is a physical difference between the
formation mechanisms of different bulge types.

\begin{figure}[t]
\begin{center}
\includegraphics[width=0.50\textwidth]{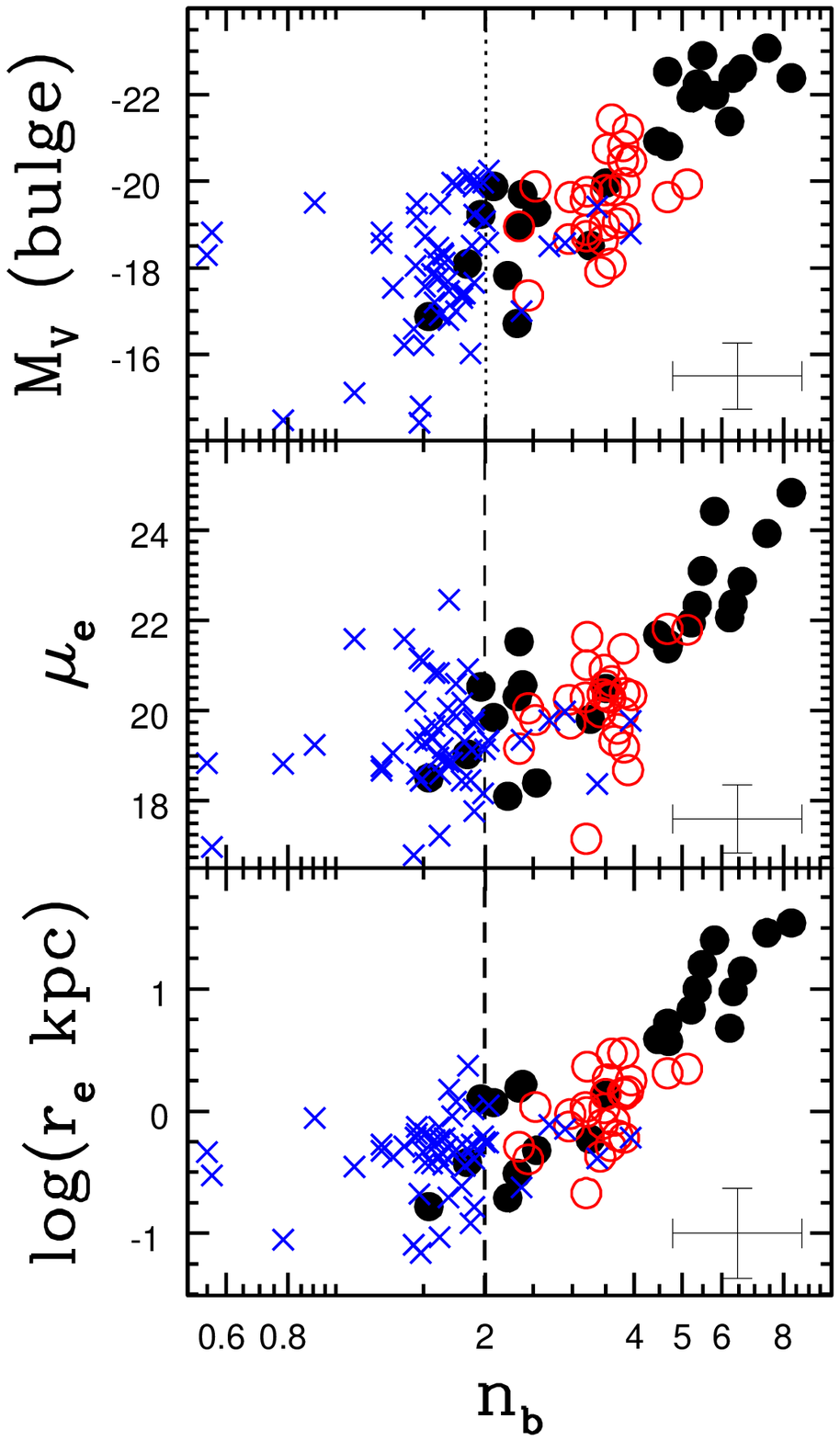}
\end{center}
\caption{Correlation of the absolute magnitude ($M_V$), half-light
  radius ($r_e$), and surface brightness at the half-light radius
  ($\mu_e$) of bulges with bulge S\'ersic index. Pseudobulges are
  represented as blue crosses, classical bulges as red circles. We
  also show elliptical galaxies from \cite{kormendy2006virgo} for
  comparison, represented by black filled circles. The average
  uncertainty of the parameters of all bulges is represented by the
  error bars in the bottom right corner of each
  panel. \label{fig:eff_param}}
\end{figure}

\begin{figure}[t]
\begin{center}
\includegraphics[width=0.48\textwidth]{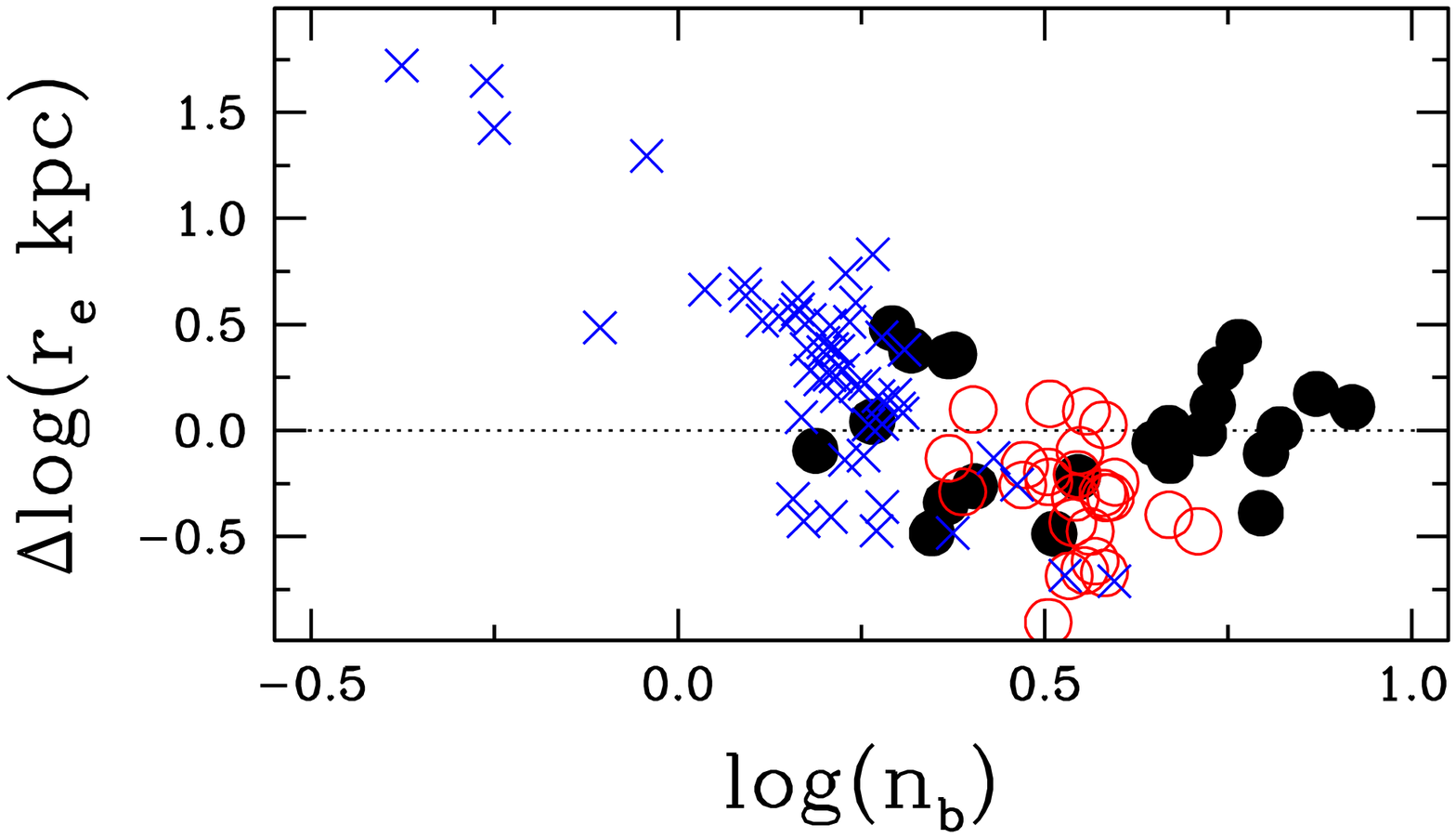}
\end{center}
\caption{Residuals of pseudobulges (crosses), classical bulges (open
  circles), and elliptical galaxies (closed circles) from a fit to the
  elliptical galaxies given by
  $\log(r_e/\mathrm{kpc})=2.9\log(n_b)-1.2$.\label{fig:resid} }
\end{figure}

\subsection{The S\'ersic Index In Bulges}
Figure~\ref{fig:eff_param} shows the correlation of bulge properties
with the shape of the surface brightness profile measured by the
S\'ersic index, $n_b$, of those bulges. We find that no classical
bulge has S\'ersic index less than two, and very few pseudobulges
($\sim 10\%$) have S\'ersic index greater than 2.  The correlations of
S\'ersic index with other structural properties ($r_e$, $\mu_e$, and
$M_v$) appear to be different for pseudobulges and classical bulges.
This suggests that the S\'ersic index may be used to distinguish
pseudobulges from classical bulges.

The top panel of Fig.~\ref{fig:eff_param} shows the correlation of
S\'ersic index with absolute $V$-band bulge magnitude for
pseudobulges, classical bulges, and elliptical galaxies.  Classical
bulges fit well within the correlation set by elliptical galaxies.
They cover a range of S\'ersic index from $n_b\sim 2-4.5$, and their
absolute magnitude covers the range $M_V\sim -17.5$ to $-21$.  Notice
that the S\'ersic index of classical bulges is normally smaller than
$n=4$, suggesting their similarity to the low-luminosity ellipticals
discussed by \cite{kormendy2006virgo}. Classical bulges, yet again,
resemble little elliptical galaxies that happen to be surrounded by a
disk.  The mean pseudobulge absolute magnitude is $\sim 1.2$
magnitudes fainter than the average classical bulge magnitude, and the
faintest bulges are all pseudobulges. Yet, pseudobulges do no have to
be faint: classical bulges are not much brighter than the brightest
pseudobulges. Our data cannot show if the brightest pseudobulges are
bright because of the nature of their stellar population, or because
of greater stellar mass.  Nonetheless, pseudobulges are not merely a
change in observed properties at low magnitude.  Finally, we note that
the overall shape of the $M_V-n$ correlation for the super-set of
systems represented in Fig.~\ref{fig:eff_param} is similar to those
found in previous studies, \citep{graham2001,balcells2003,dejong2004}.

The bottom two panels of Fig.~\ref{fig:eff_param} show the correlation
of S\'ersic index with the half-light quantities (radius containing
half the light, $r_e$, and surface brightness at that radius, $\mu_e$)
of pseudobulges, classical, bulges, and elliptical galaxies. The
results are similar to those in the $M_V-n$ plane. Classical bulges
exist within the bounds set by elliptical galaxies, more specifically
by low-luminosity ellipticals.  Pseudobulges, on the other hand,
populate the small $n$ extreme of this correlation. The smallest
objects, in radius, are pseudobulges.  Yet, pseudobulges in general
are not that much smaller on average than classical bulges and
low-luminosity ellipticals: the mean effective radius of pseudobulges
lies within the range of effective radii spanned by the classical
bulges and elliptical galaxies.
\begin{figure}[t]
\begin{center}
\includegraphics[width=0.48\textwidth]{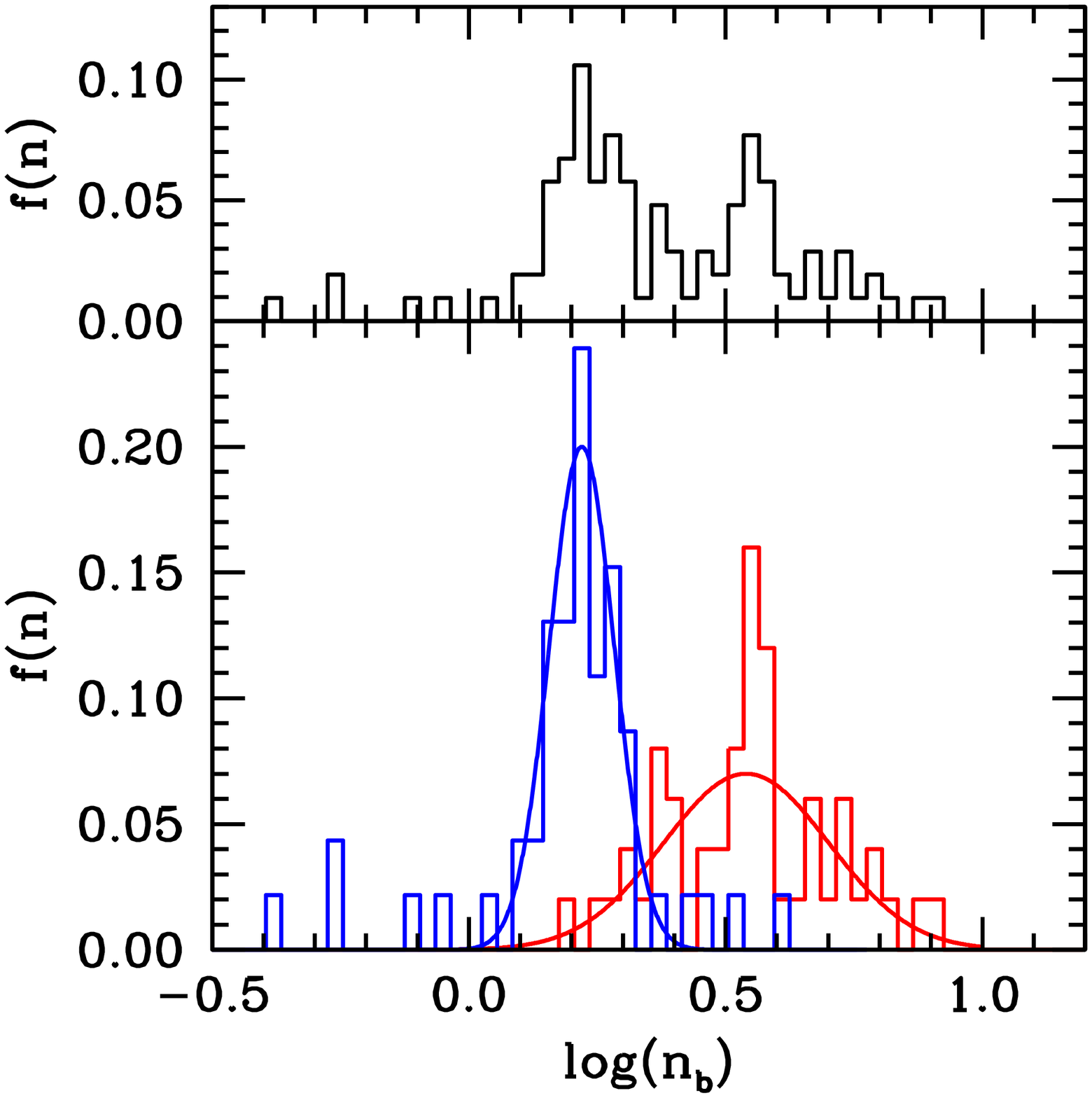}
\end{center}
\caption{Histograms of bulge S\'ersic index, $\log(n_b)$, by bulge
  type. The top panel shows the histogram of all objects. The bottom
  panel shows pseudobulges (blue lines with mode $\log(n_b)\sim0.22$)
  and classical bulges + elliptical galaxies (red lines, mode
  $\log(n_b)\sim0.54$). We also overplot Gaussian distributions that
  fit the histograms. Histograms are normalized by the total
  number of galaxies in each group, so that each bin reflects the
  frequency of S\'ersic index within that object
  class. \label{fig:dist}}
\end{figure}

To more strongly emphasize the difference between pseudobulges from
classical bulges and ellipticals (hot stellar systems), we show the
residuals of all galaxies to the \cite{caon94} relation $r_e\propto
\log(n_b)$ in Fig.~\ref{fig:resid}. We fit this correlation to the
elliptical galaxies only, and the resulting relation is $\log(r_e/
\mathrm{kpc}) = 2.9 \log(n_b) - 1.2$.  E-type galaxies and classical
bulges have similar amounts of scatter. Pseudobulges show a marked
difference and systematically deviate further from the correlation as
$n_b$ gets smaller whereas the scatter among classical bulges is
contained within the region occupied by ellipticals. Thus it appears
that pseudobulges do not relate $n_b$ to $r_e$ in the same way as
elliptical galaxies and classical bulges do. It is not clear from the
bottom panel of Fig.~\ref{fig:eff_param} whether S\'ersic index
correlates at all with effective radius for pseudobulges.

Inspection of Fig.~\ref{fig:eff_param} shows that pseudobulges and
classical bulges have separate distributions of bulge S\'ersic index;
the value that distinguishes the two types of bulges appears to be
$n_b\sim 2.$ Pseudobulges have S\'ersic index smaller than $n_b \sim
2$; classical bulges have larger values than $n_b \sim 2$. We show
this explicitly in Fig.~\ref{fig:dist} which shows the distributions
of $\log(n_b)$ for pseudobulges (blue lines) and the superset of
classical bulges and elliptical galaxies (red lines), both binned to
$\delta \log(n_b)=0.03$. We also show a histogram of all objects
(pseudobulges, classical bulges, and elliptical galaxies) counted
together.  Each histogram is normalized by the number of objects in
that group, so that the counts reflect the frequency of S\'ersic index
within each class. Before discussing these distributions any further,
we must strongly qualify this result. The sample studied in this paper
is in no way complete.  Thus, the relative abundance of pseudobulges
and classical bulges cannot be determined from this
distribution. However, we wish to study the distribution of S\'ersic
indices in different bulge types. We do the best we can with the data
we have at present; when relevant, we attempt to point out how our
sample selection may bias results. We note that distributions of bulge
S\'ersic indices in volume limited samples have been shown to be
bimodal \citep{allen2006}. We present any numbers derived from
analysis of these distributions as preliminary.

The top panel of Fig.~\ref{fig:dist} shows that the distribution of
S\'ersic indices are clearly bimodal. This bimodality is completely
coincident with the dichotomy in bulge morphology. The average values
of bulges S\'ersic index for pseudobulges and classical bulges are
$1.69\pm0.59$ and $3.49\pm0.60$, respectively, and the average of the
superset of classical bulges and elliptical galaxies is $3.78\pm
1.5$. The range of Hubble types chosen for this sample (S0-Sc)
certainly biases the distribution of S\'ersic index to higher values
since the S\'ersic index has been found to get smaller at later Hubble
types \citep{graham2001,macarthur2003}. It is worth noting that the
average uncertainty of a bulge (pseudo- or classical) is comparable to
the standard deviation of either of these distributions, $\langle
\Delta n_b \rangle = 0.60$. Inspection of Fig.~\ref{fig:eff_param}
shows that neither $\log(r_e)$ nor $\mu_e$ show a bimodal
distribution.

We fit a Gaussian to each histogram in Fig.~\ref{fig:dist} and
solve for the S\'ersic index where the frequencies are
equal. Because our sample is not volume limited, we weight the
distributions such that the number of galaxies earlier than Hubble
type Sc with pseudobulges is equal to the number of galaxies with
classical bulges. We also solve for the case with 1/3 pseudobulges, and
finally 2/3 pseudobulges. We find the critical S\'ersic index to be
$n_{crit} = 2.2\pm 0.1$.
\begin{figure}[t]
\begin{center}
\includegraphics[width=0.44\textwidth]{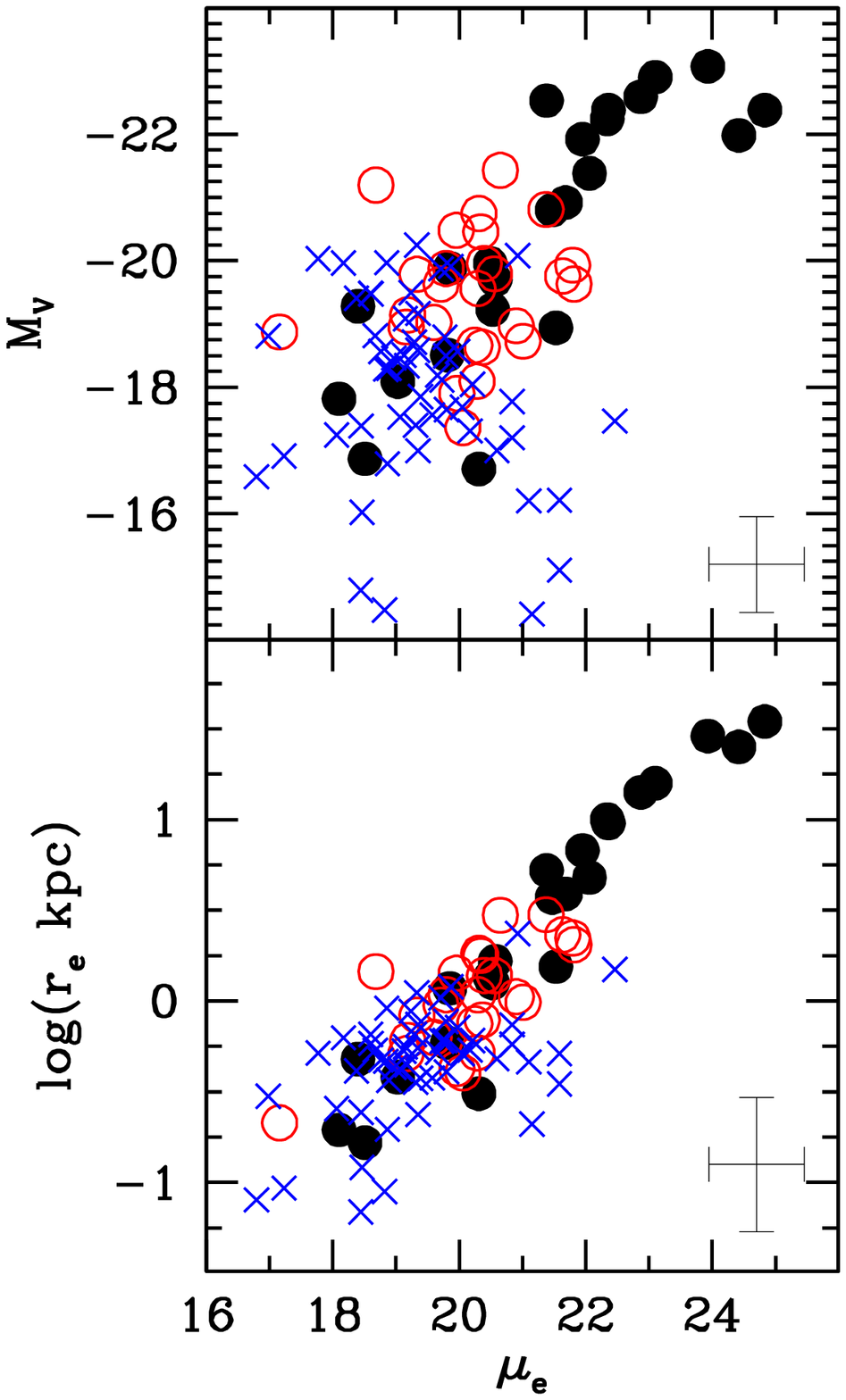}
\end{center}
\caption{Two photometric projections of the fundamental plane. All
  symbols are the same as figure \ref{fig:eff_param}.\label{fig:fund}
}
\end{figure}

\begin{figure*}[t]
\begin{center}
\includegraphics[width=0.99\textwidth]{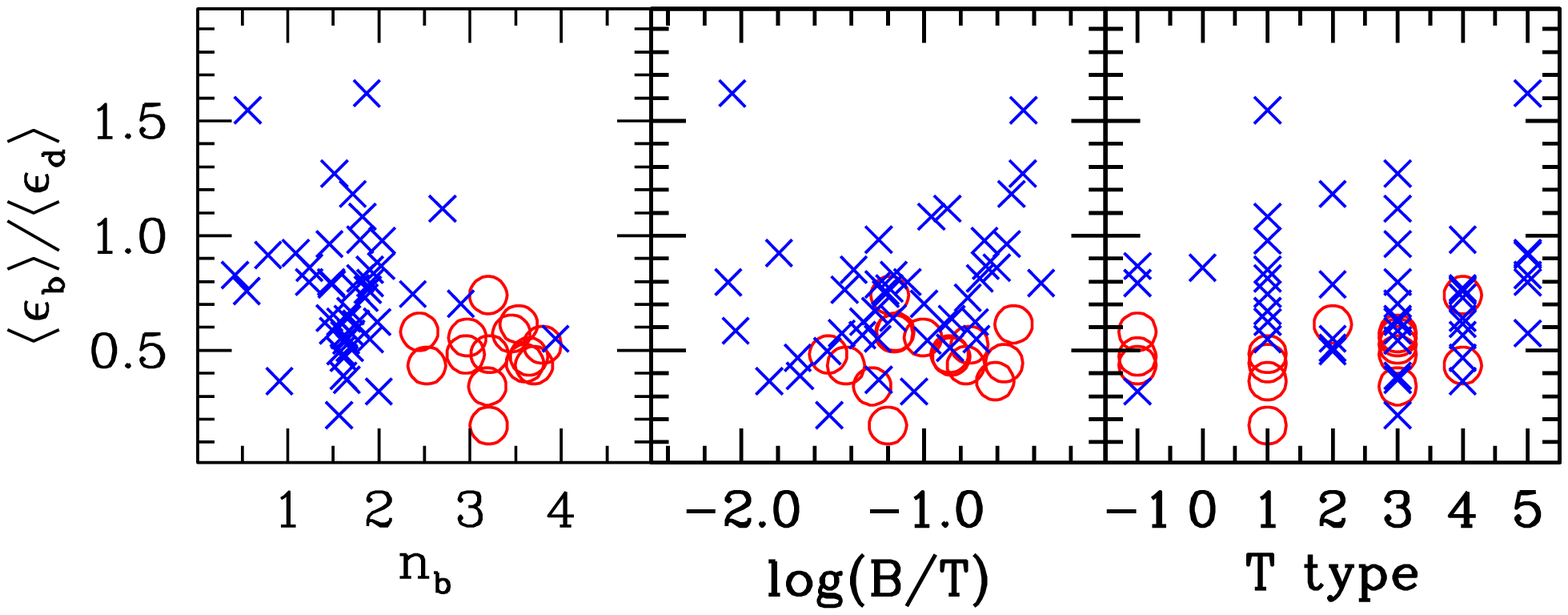}
\end{center}
\caption{Ratio of average ellipticity of the bulge to that of the
  disk. As usual, pseudobulges are represented by blue crosses and
  classical bulges by open red circles. This ratio is plotted against
  S\'ersic index of the bulge (left), bulge-to-total luminosity ratio
  (middle), and Hubble type (right). \label{fig:ellip}}
\end{figure*}
Of the 52 pseudobulges, 5 have S\'ersic index significantly above two
(two pseudobulges have $n_b\sim 2$). The interested reader can inspect
their surface brightness profile in the appendix, which shows fits to
all surface brightness profiles. We take a few lines to discuss these
galaxies here. NGC~4314 is an SBa galaxy with $n_b=2.37\pm 0.78$; this
fit appears good, although a significant amount of the profile is not
included due to the outer bar and nuclear ring. NGC~4258 is an oval
galaxy, with $n_b=2.69\pm 0.48$. The outer oval ring affects the
surface brightness profile of the outer disk, and thus the fit covers
a narrow range. However, this is unlikely to affect the fit of the
bulge too much, especially given the relatively low uncertainty in
S\'ersic index in this fit. NGC~3627 is an Sb galaxy with $n_b=2.90\pm
0.83$. Its bar is not easily detectable, and thus not removed from the
fit, as we try not to remove any unnecessary points. NGC~3642 is an Sb
galaxy with $n_b=3.37\pm0.61$, the profile does not go very deep, and
may be allowing $n_b$ to be artificially high. Finally, NGC~4826 has
the largest S\'ersic index of any bulge classified as a pseudobulge,
and the fit appears quite good, unlike the others. We take these
galaxies as exceptions rather than the rule, future studies comparing
alternative methods of pseudobulge detection (e.g.\ kinematics) will
help shed light on their true nature. We remind the reader again that
we are not classifying systems as composites, and these may have an
effect. If for example a pseudobulge is embedded in a classical bulge
(or vice versa), we might visually identify the pseudobulge even
though the stellar light distribution is set by the classical
bulge. Still, the dividing line of $n_b \sim 2$ is 90\% successful at
identifying pseudobulges, and thus is a good detector of pseudobulges.

\subsection{Fundamental Plane Projections}

Figure~\ref{fig:fund} shows the correlations of $\mu_e-r_e$ and
$M_v-\mu_e$, two projections of the fundamental plane
\citep{djorgovski1987,dressler1987,faber1989}. The top correlation
shows the magnitude versus surface brightness plane, and the bottom
panel shows the $\mu_e-r_e$ relation \citep{k77}.  While the $M_V$
versus $\mu_e$ plot has a lot of scatter, especially in the
pseudobulges, the radius-surface brightness plane shows significantly
less scatter. Yet in both of these fundamental plane projections,
those bulges which are further from the correlations established by
the elliptical galaxies are pseudobulges. For those pseudobulges that
deviate from the fundamental plane correlation in Fig.~\ref{fig:fund}
that deviation is towards lower densities.

This behavior has been noticed in the past.  \cite{carollo1999} finds
that exponential bulges are systematically lower in effective surface
brightness than those better fit by an $r^{1/4}$-profile. We note that
this could be an effect of only fitting either $r^{1/4}$ or
exponentials rather than a S\'ersic function to the bulge profiles.
If a bulge is not completely exponential then it may force other
parameters to compensate for the more restrictive parameterization of
the profile shape. \cite{falcon2002} find that bulges which deviate
from the edge-on projection of the fundamental plane are found in late
type (Sbc) galaxies. However, at least with the galaxies in our
sample, we cannot say unambiguously that this is a function of
differing bulge formation, because those bulges in Fig.~\ref{fig:fund}
that deviate significantly from the correlations defined by the
elliptical galaxies are also very low $B/T$. Thus, it could be that
the potential of the bulge is more affected by the outer disk.

In fact, when looking at the pseudobulges in Fig.~\ref{fig:fund}
alone, one would not infer the presence of a strong correlation of
either magnitude or effective radius with surface brightness. This is
another indication that pseudobulges and classical bulges are
different classes of objects.
\begin{figure*}[t]
\begin{center}
\includegraphics[width=0.99\textwidth]{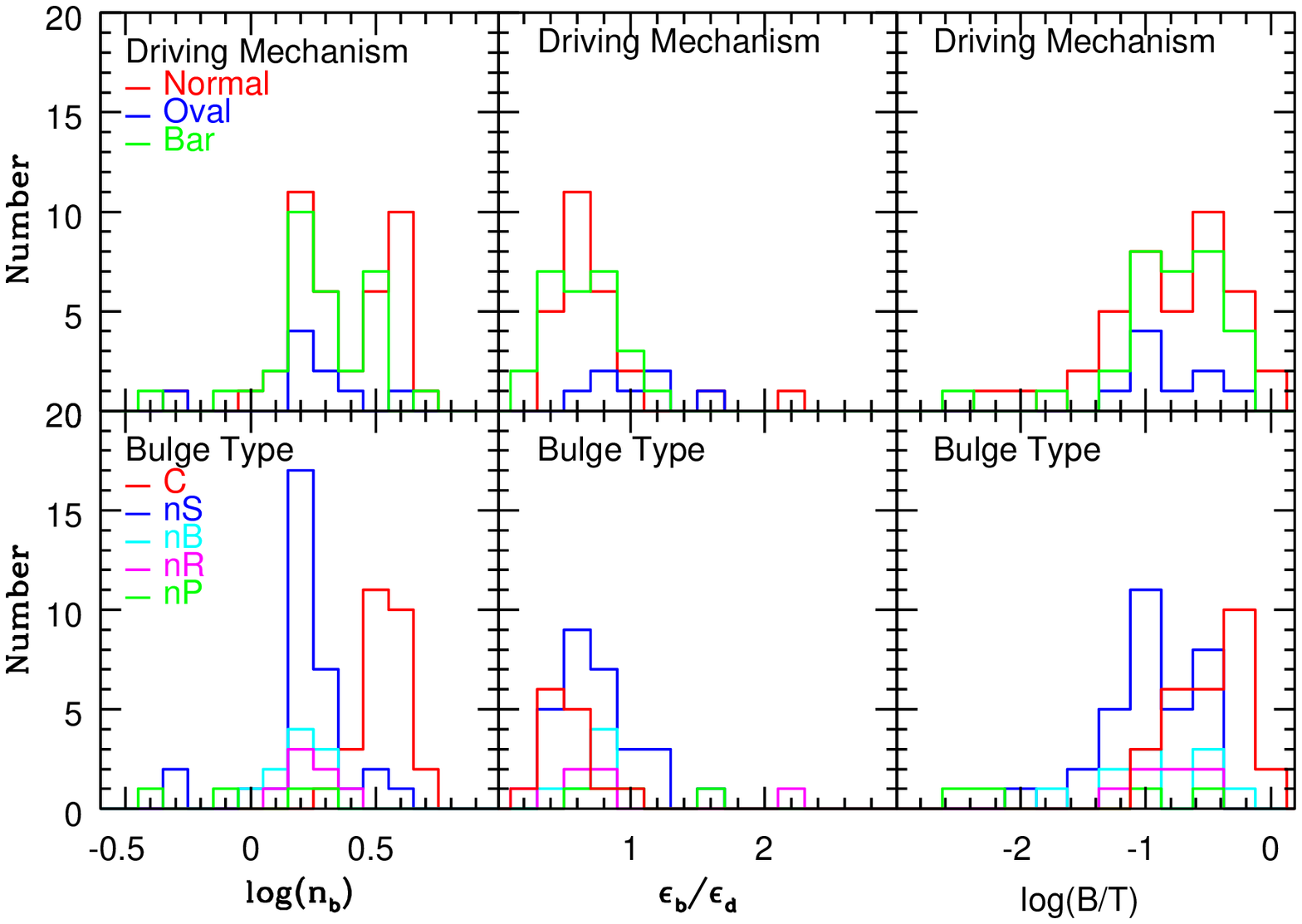}
\end{center}
\caption{In the bottom panels, we show the distribution of o S\'ersic
  index (bottom left), ratio of ellipticities (bottom middle), and
  $B/T$ (bottom right) for different bulge morphologies. The bulge
  types are denoted by nS=nuclear spiral, nB=nuclear bar, nR=nuclear
  Ring, nP=nuclear patchiness, and C=classical bulge. We also show the
  comparison of the presence of driving mechanism (N=no driver,
  O=oval, and B=bar) to the S\'ersic index (top left), ratio of
  ellipticities (top middle), and $B/T$ (top
  right). \label{fig:bmorph} }
\end{figure*}

\section{Flattening Of Classical Bulges And Pseudobulges}

There has been very little work on the distribution of flattenings of
bulges albeit the structures present in pseudobulges (e.g.\ nuclear
spirals or nuclear bars) suggest that pseudobulges should have higher
angular momentum and thus be flat stellar systems. However, the end
products of secular evolution need not be flat \citep{k93}. Bar
buckling and unstable disks can both drive stars higher above the
plane of the disks \citep{pfenniger1990,friedli1999}, thus creating a
pseudobulge that is less flat than its associated outer
disk. Nonetheless, the data of \cite{kent1985} show that many bulges
have median flattenings that are similar or greater than the median of
the outer disk, and that flat bulges are more common in late-type
galaxies. \cite{fathi2003} find a similar result that
$\epsilon_{bulge}/\epsilon_{disk}>0.9$ in 36\% of S0-Sb galaxies and
51\% Sbc-Sm galaxies (where $\epsilon=1-b/a$). However, note that
\cite{mollenhoff2001} do not find this result, they find very few
bulges are as flat as disks in ground based $JHK$ imaging.

\cite{kk04} include the flattening of bulges (as manifest through the
ratio of bulge ellipticity to that of the disk) in their list of
preliminary criteria for identifying pseudobulges. We can test this
hypothesis with our sample.  Figure~\ref{fig:ellip} shows the ratio of
mean ellipticity of the bulge to that of the outer disk ($\langle
\epsilon_b\rangle/\langle\epsilon_d\rangle$) for galaxies in this
study. We do not include galaxies with average disk ellipticity less
than 0.2, as face-on projections of galaxies do not allow the
flattening to be determined. Also, galaxies with $B/T > 0.5$ are
removed; if the bulge dominates the entire potential it may set the
shape of the disk and therefore affect the disk's ellipticity making
it more like its own. To calculate the average ellipticity of the
profile we use only the data points that are also included in the fit
(filled circles in Figs.~\ref{fig:egprof} and in the Appendix). It is
a matter of interpretation as to what features are a part of
pseudobulges, especially in light of the fact that some nuclei may be
formed secularly along with the pseudobulge. However, as a matter of
consistency we choose to focus only on those isophotes we call the
``bulge'' from the bulge-disk decompositions. The boundary between the
bulge and disk is chosen as the radius at which the surface brightness
of the S\'ersic function equals the surface brightness of the
exponential disk in the decomposition. Yet, in many galaxies
contamination from the disk artificially raises the average
ellipticity of the bulge. This contamination is evident in the
ellipticity profile. Thus, we choose to average the bulge over a
region in which there is little contamination present. Those radii are
given in Table~2.

The leftmost panel in Fig.~\ref{fig:ellip} compares $\langle
\epsilon_b\rangle/\langle\epsilon_d\rangle$ to bulge S\'ersic index
for pseudobulges and classical bulges. It is quite evident that
classical bulges, in our sample, are not as flat as pseudobulges. The
average ratio of ellipticities of pseudobulges is $0.79\pm 0.1$,
whereas the average ratio for classical bulges is $0.49\pm 0.14$. In
fact, the flattest classical bulge ($\langle
\epsilon_b\rangle/\langle\epsilon_d\rangle=0.75$) is less flat than
the average pseudobulge.
\begin{figure}[t]
\begin{center}
\includegraphics[width=8cm]{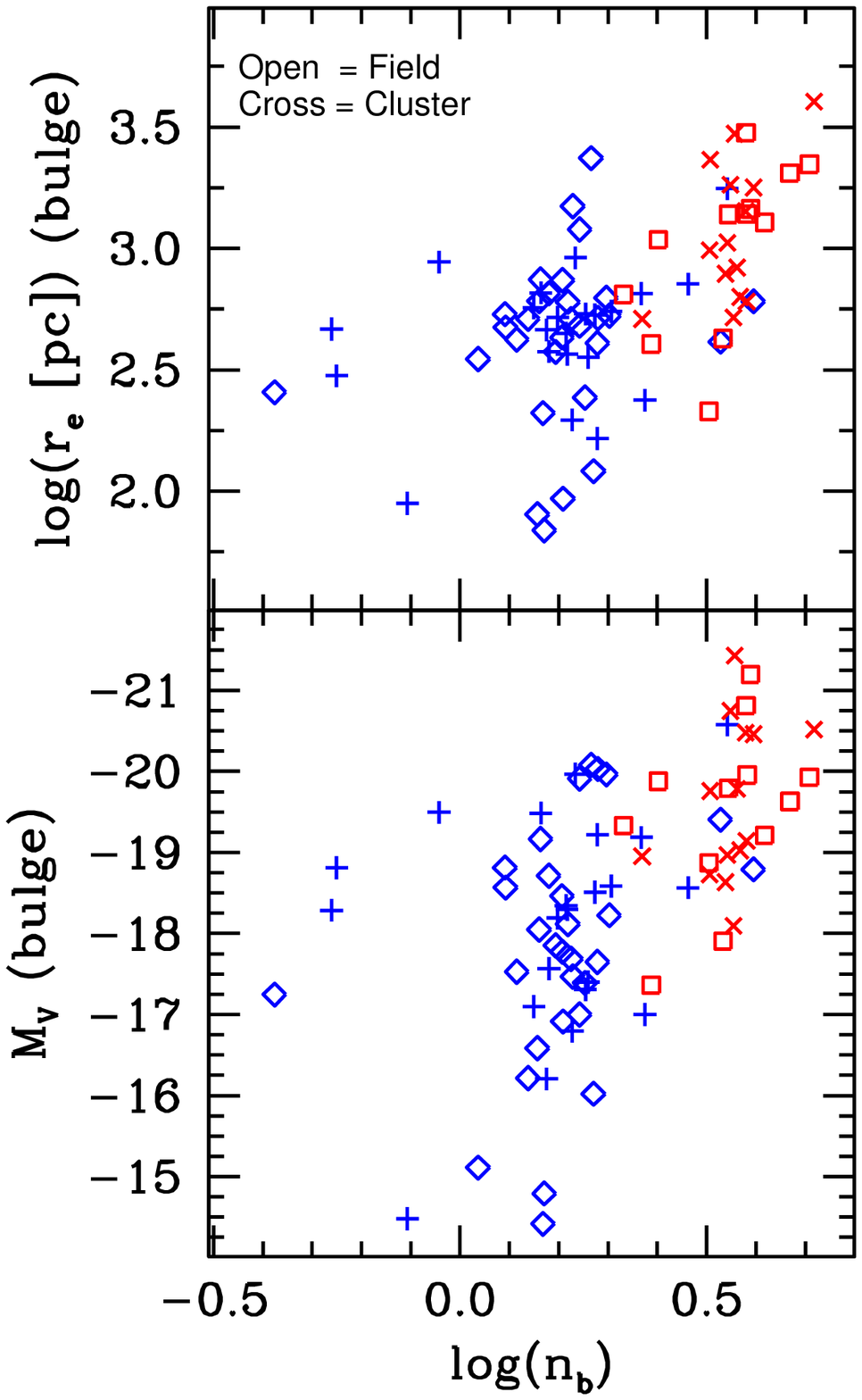}
\end{center}
\caption{Structural parameters of bulges as a function of the environment. 
Field glaxies with classical bulges are
marked by open squares, field pseudobulges with open diamonds. Cluster
galaxies with classical bulges are marked by X's, cluster pseudobulge
galaxies by crosses. All classical bulges are marked by red symbols, pseudobulges by blue symbols.
\label{fig:env_n} }
\end{figure}

\begin{figure}[t]
\begin{center}
\includegraphics[width=8cm]{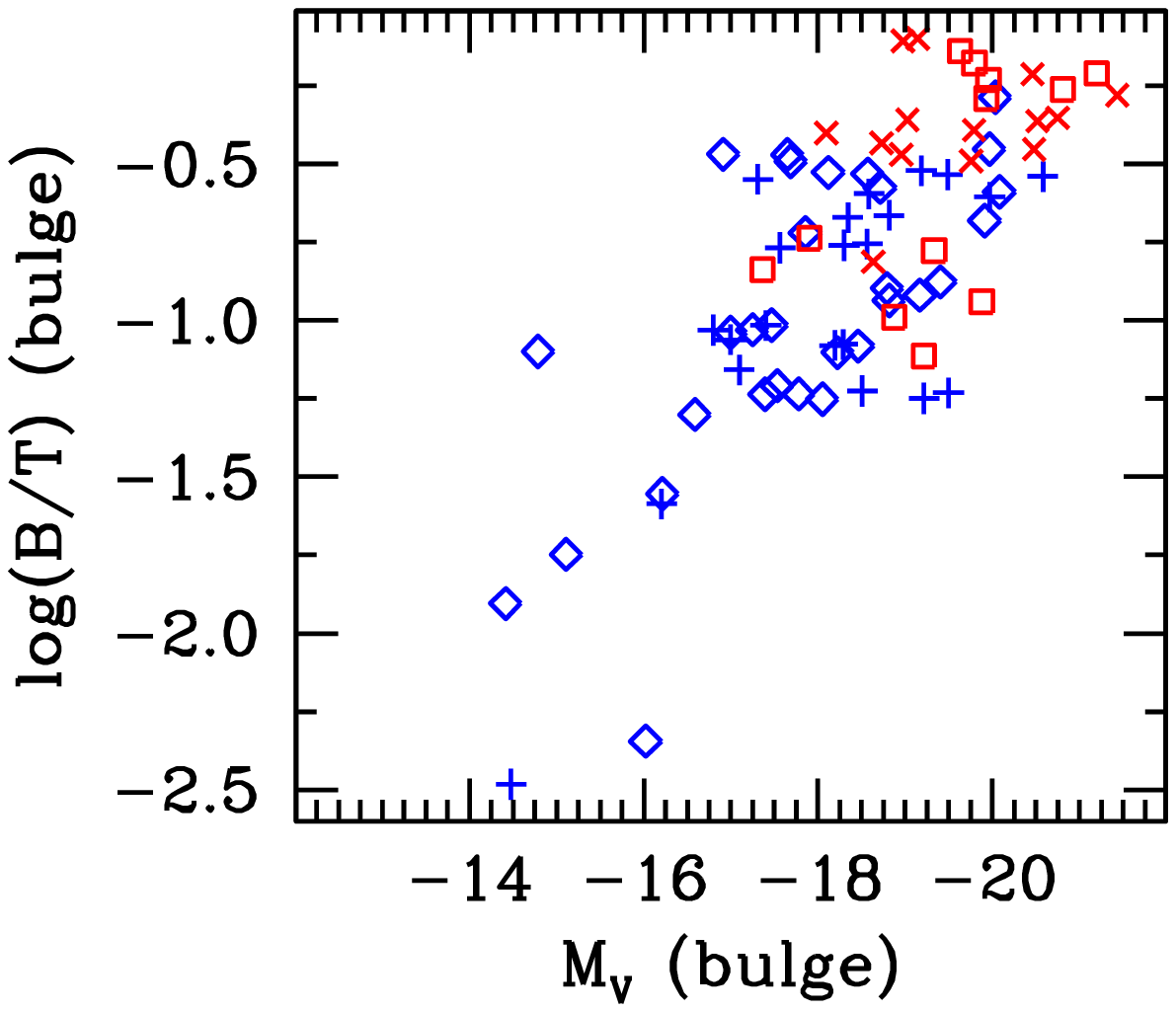}
\end{center}
\caption{As in fig.~\ref{fig:env_n}, we distinguish those those
  galaxies that exist in cluster (or group) environments (crossed
  symbols) from those that exist in the field (open symbols). Symbol
  color reflects bulge type, as before. Bulge-to-total ratio is
  plotted against absolute magnitude of the bulge.\label{fig:env_bt} }
\end{figure}

The middle and rights panels of Fig.~\ref{fig:ellip} compare this
ratio to the bulge-to-total ratio and Hubble type,
respectively. Classical bulges show no obvious trend of $\langle
\epsilon_b\rangle/\langle\epsilon_d\rangle$ with $B/T$ or Hubble
type. They are tightly clustered about the mean value $\sim0.6\pm
0.1$. Pseudobulges however show a slight correlation of $\langle
\epsilon_b\rangle/\langle\epsilon_d\rangle$ with $B/T$. This
correlation indicates that more prominent (in $B/T$) pseudobulges are
slightly flatter. However, the trend is weak, and has large
scatter. Finally, we find a similar result to \cite{k93} and
\cite{fathi2003}, namely that flatter bulges are more frequent in
later-type galaxies. Furthermore, this trend only exists for
pseudobulges.

Invariably the flattest bulges are pseudobulges. However, we do not
wish to overstate this result. It is possible that asymmetric dust
extinction leads to higher apparent flattening in the V band. Future
work at near-to-mid IR wavelengths is likely to provide less ambiguous
results.

\section{Pseudobulges As A Class Of Objects}

In this paper we identify pseudobulges as having any, but not
necessarily all, of several structures (nuclear bars, rings, and
ovals). We treat bulges with these structures as a group. The
motivation for doing so is that all these phenomena are similar to
properties that are commonly associated with high specific angular
momentum systems. At least in the sense that they are the complement
of classical bulges they can be treated as a group. However, here we
have to ask the following question: do thusly identified pseudobulges
act as a single class of objects or do significant differences exist
among the objects we are identifying as pseudobulges?

Figure~\ref{fig:bmorph} suggests an answer to this question.  The left
panels in the figure show the distribution of bulge properties for
bulges that have smooth isophotes (C=classical), bulges with nuclear
spirals (nS), bulges with nuclear bars (nB), bulges with nuclear rings
(nR), and bulges with a chaotic nuclear patchiness that resembles late
type galaxies (nP). See Fig.~\ref{fig:bulgeid} for example images of
each of these features. In S\'ersic index, ellipticity ratio (bulge to
disk), and bulge-to-total ratio there is no significant difference
among the types of morphologies we call a pseudobulge.

Bulges with chaotic nuclear-disk-like patchiness (nP) seem to have
smaller S\'ersic index, and are more flat. It is unclear why this
might be. This may be driven by them having smaller $B/T$ and thus
them being more affected by the disk potential. We also note that
pseudobulges with $n_b>2$ are almost all in bulges with nuclear
spirals. It is possible that these bulges are not truly a spiral,
instead some other phenomenon, like contamination of the light by the
outer disk, is causing us to identify them as pseudobulges. However,
in each parameter the distinction appears to be between classical
bulges and the rest of the objects, rather than among the objects we
call pseudobulges.

We also show S\'ersic index, ellipticity ratio (bulge to disk), and
bulge-to-total ratio as a function of secular driving mechanism. We
separate galaxies as having no driving mechanism, an oval disk, or a
bar. Notice once again that we do not set grand design spiral as a
class of object. There seems to be little differences in the averages,
except ovaled galaxies have smaller S\'ersic index and flatter
bulges. The phenomenon of flat bulges does not appear to require the
presence of a bar. Thus it is not likely that all flat bulges are
small bulges stretched by a bar potential making them flat. Also
notice that the distributions of parameters for barred, oval, and
normal spirals is roughly the same. Thus it does not seem likely that
our method of removing the bar is artificially changing the bulge-disk
decomposition parameters.

\section{Environmental Effects}

Secular evolution is not the only theory for building bulges that look
like disks. Other possibilities such as extremely gas rich accretion
events, distant gravitational encounters, or gravitational
interactions with a cluster potential could all drive gas to the
center of galaxies to increase $B/T$, as found by
\cite{mastropietro2005}. \cite{kannapan2004} find that blue bulges are
statistically more likely to have companion galaxies within
100~kpc. Further, counter-rotating gas is frequently observed in
spiral galaxies, and is taken as a sign of the galaxy having accreted
galaxies in their past.  Yet, pseudobulges are much more common in
late-type galaxies, and the well known morphology-density relation
\citep{dressler1980} shows that late-type galaxies are not found in
dense environments. Is it possible that pseudobulges in the field are
formed through different mechanisms than those in cluster
environments? We can not know the answer to this for certain, however
we can look for signatures for such effects in our data.

In Fig.~\ref{fig:env_n} we show our primary result again, namely that
pseudobulges have a lower S\'ersic index than classical bulges, but
here we also mark environment. Field glaxies with classical bulges are
marked by open squares, field pseudobulges with open diamonds. Cluster
galaxies with classical bulges are marked by X's, cluster pseudobulge
galaxies by crosses.  There seems to be no significant differences in
the structure of the surface brightness profile between the
pseudobulges that reside in the field galaxies and the ones in cluster
galaxies. Cluster pseudobulges are not higher mass nor systematically
different in radial size than field pseudobulges. Nor are their
surface brightness profiles preferentially steeper in clusters.

The same holds for $B/T$: there is no substantial difference between
pseudobulges in the field versus clustered environments. If
pseudobulge formation was driven primarily by tidal encounters with
distant galaxies, one would expect that this effect should be enhanced
in cluster environments, where such encounters are more frequent. This
would result in more massive pseudobulges existing in cluster
environments. Our sample does not seem to indicate that pseudobulges
are any more luminous in cluster environments than in the field. We
feel that this supports the notion that externally driven disk
evolution is not likely the dominant affect in driving pseudobulge
formation.

Our results in no way rule out the possibility that pseudobulges are
formed by gas-rich minor mergers. If we take an example of merging our
Galaxy with one of the Magellanic clouds, and if this is done $\sim$1
Gyr, when the gas fractions were much higher, it is entirely plausible
that the result could look similar to what we call a
pseudobulge. However, it is not certain if such a system would still
be actively forming stars today, or if pseudobulge formation through
minor mergers could only happen in extreme cases (e.g.\ prograde
collisions at low inclination). Also, it is not clear how such
accretion-driven formation of pseudobulges could maintain the
bulge-disk correlations discussed in the previous sections.

\section{Summary \& Discussion}

The main result of this paper is that bulge S\'ersic index, $n_b$, is
bimodally distributed in intermediate type galaxies where both
classical bulges and pseudobulges exist. A value of $n_b\simeq 2$
marks the boundary for separating morphologically-identified
pseudobulges from classical bulges. Below $n_b=2$ no classical bulges
are found, and above it very few pseudobulges are found. We also find
that galaxies which are identified as pseudobulges, using either bulge
morphology or bulge S\'ersic index, are flatter than classical
bulges. Thus suggesting that on average these systems are more
disk-like in both their morphology and shape than are classical
bulges. Pseudobulges exist that are as round as classical bulges, yet,
invariably the flattest bulges are all pseudobulges. In both S\'ersic
index and flattening ratio, our results show a homogeneity in
classical bulge properties, and a greater dispersion in pseudobulge
properties. That is to say, classical bulges, in our sample, do not
have S\'ersic index less than two, nor do they have high ratios of
bulge-to-disk ellipticity. Pseudobulges, on average, have lower $B/T$
than classical bulges, however we find pseudobulges with $B/T$
extending to $\sim 0.35$.

We find that the half-light radius of pseudobulges is well correlated
with the scale length of the outer disk, while the scale length of
classical bulges is not correlated with that of the disk. This is
consistent with the interpretation that $r_e \propto h$ is due to a
secular formation of pseudobulges.  Also, the fact that the scale
length of the disk is correlated with the size of its pseudobulge but
not correlated with the size of classical bulges is a strong
indication that there indeed is a physical difference between the
formation mechanisms of these different bulge types.

In photometric fundamental plane projections, pseudobulges populate
and extend the low luminosity end of the range occupied by classical
bulges and elliptical galaxies. Pseudobulges deviate more from the
relations set by the elliptical galaxies than classical bulges, and
preferably towards lower density. In fact, it does not seem that on
their own pseudobulges would establish any of these correlations,
especially $M_V - \mu_e$.

In all correlations investigated in this work we find that the
fundamental distinction is between classical bulges and pseudobulges.
We do not find significant differences within the class of
morphologies we identify as pseudobulges (nuclear spirals, nuclear
bars, nuclear rings, and chaotic nuclear patchiness).

Is it possible that identifying pseudobulges visually using nuclear
morphology is subject to an inherit flaw? That being that these
systems are by definition bulge-disk systems, and thus is it possible
that classical bulges coexist over large radius with the central parts
of a disk. Pseudobulges in S0 galaxies exemplify this concern. In many
ways elliptical galaxies and S0s are thought to be the extremes of a
continuum of properties.  Embedded disks are known to exist in
elliptical galaxies (see, e.g.\ \citealp{SB95}). Would we call such a
system where the embedded disk had a nuclear bar a pseudobulge? Most
likely not. However, in S0s, which by definition have lower $B/T$, we
are inclined to do so. Clearly, if there is no sharp change in
properties along the E-S0 continuum, there should be some composite
systems with secularly formed structure in the disk and a classical
(i.e.\ kinematically hot) bulge. (Note, though, that many S0s resemble
defunct later type disk galaxies much more than they do resemble
elliptical galaxies as suggested by \citealp{vdB76}). This may be the
case in NGC~2950, yet it does not appear to be the
rule. \cite{erwin2004composites} show that many bulges have more
complicated dynamical profiles, with both hot an cold components. Yet,
pseudobulges show a remarkable similarity to disk stellar populations,
ISM, and star formation rates. These similarities have been shown in
the flatness of stellar population gradients \citep{peletier1996}, and
the similarity of CO profiles to optical light
\citep{regan2001bima}. \cite{helfer2003} show that many CO gas maps of
bulge-disk galaxies have holes in their center, but also many CO
profiles rise steeply to their center. A quick comparison of with our
sample shows that those holes are found in classical bulges (e.g.\
NGC~2841).  \cite{fisher2006} shows that the nuclear morphology of
galaxies (as used here) predicts the shape of the 3.6-8.0 micron color
profile: disk-like bulges have flat color profiles, and E-like bulges
have holes in 8.0 micron emission. If we were merely mistakenly
identifying disks superimposed on bulges as physically different
pseudobulges, we'd expect to find larger $B/T$ for such systems as
compared to the bulges where we do not see evidence for the presence
of a disk which we call classical in this work. This is because of the
added light of the disk onto a fraction of bulges drawn from the same
underlying distribution. However, we do not find this to be the case.
Fig.~\ref{fig:bthist} shows that pseudobulges are on average smaller
than classical bulges.
\begin{figure}[t]
\begin{center}
\includegraphics[width=8cm]{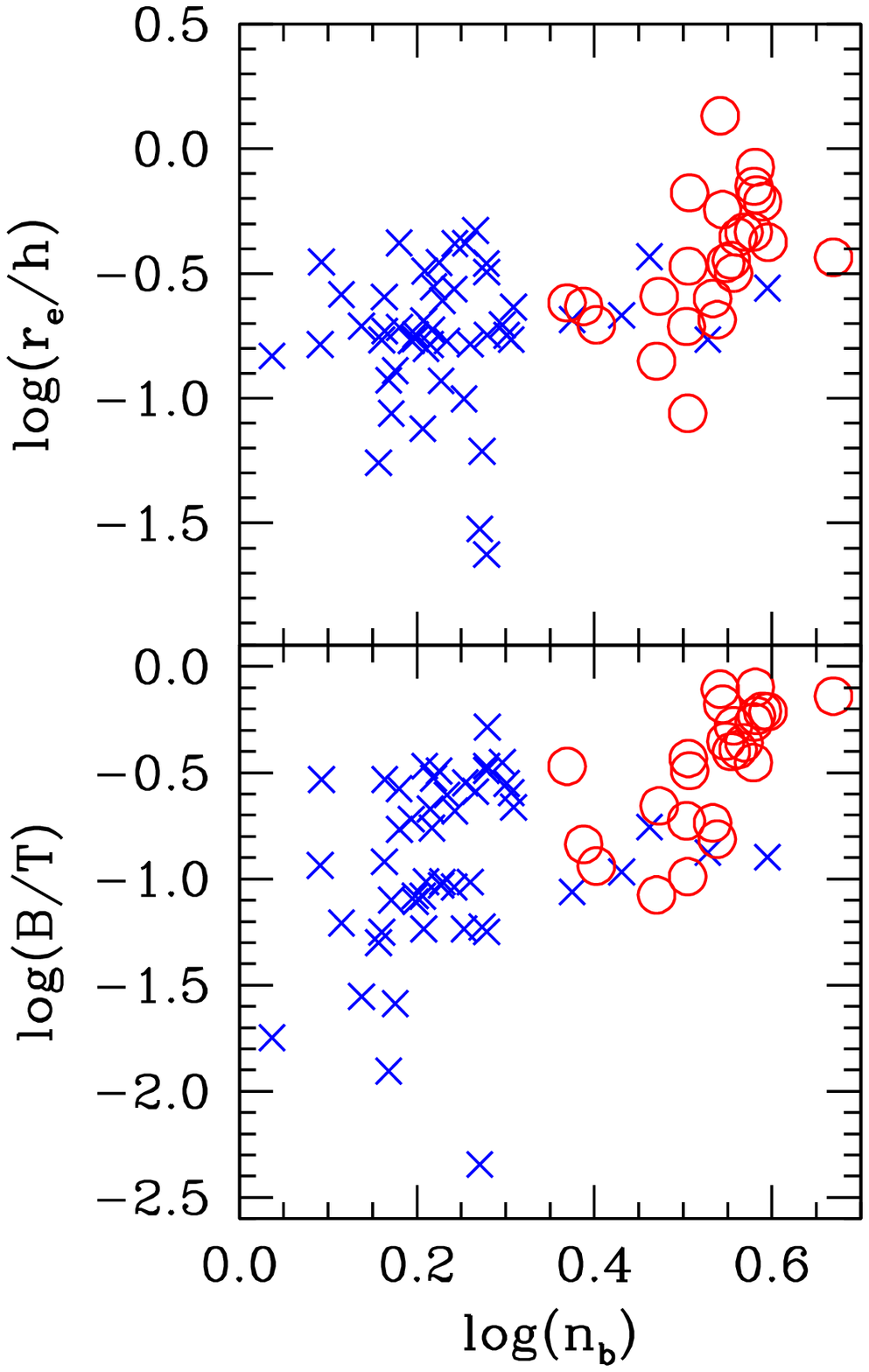}
\end{center}
\caption{Bulge-to-total ratio (lower panel) and ratio of the bulge
  half-light radius to the scale length of the disk (top panel)
  plotted against the logarithm of the bulge S\'ersic index. As
  before, classical bulges are represented by open red circles and
  pseudobulges by blue crosses.\label{fig:btn} }
\end{figure}

It is not clear whether pseudobulges participate in the various
correlations presented in this paper. If we look at the pseudobulges
alone, they only show a convincing correlation in the Kormendy
relation ($\mu_e-r_e$). Their distribution better resembles scatter
diagrams in the $M_v-\mu_e$, $\mu_e-n$, and $r_e-n$ planes. However,
it would seem to be an odd coincidence that in every parameter
combination pseudobulges just happen to fall in the right location to
extend the correlations set by the classical bulges and the elliptical
galaxies. Putting all this together implies that we do not really
understand the details of pseudobulge formation very well. It is also
possible that the decomposition of pseudobulges may not be an
appropriate procedure. Bulge-disk decomposition assumes that the
structures are distinct (and that light extends to radius of
infinity). If we accept that pseudobulges are formed through disk
phenomena, then such a bulge-disk decomposition may not be an adequate
description of those systems. In reality they might simply be a
component of galactic disk. Also, parameters derived from the S\'ersic
function are coupled \citep{graham1997}. Thus, treating pseudobulges
in the same way as classical bulges may artificially force them to
extend some photometric correlations.

Concerning the six bulges with disk-like morphology near their centers
that have $n_b>2$, two have $\langle
\epsilon_b\rangle/\langle\epsilon_d\rangle > 0.75$ (the maximum of
classical bulges), and NGC~4314 has a very prominent star forming
nuclear ring, making this bulge a strong candidate for being a
pseudobulge. This galaxy might be an example for a composite
system. Their existence certainly underscores the value of having as
much information as possible when diagnosing bulge types. Future work
may prove illuminating. For example, to what extent should we trust
bulge morphology as an indicator of secular evolution? It is possible
for the human eye to mistake merely the presence of dust for spiral
structure?

We also find that the ratio of scale lengths, $r_e/h$, in pseudobulges
is more tightly correlated and closer to those values reported by
other authors (e.g.~\citealp{macarthur2003}) than in classical
bulges. However, the range in values of $r_e/h$ for classical bulges
and pseudobulges are similar; therefore this does not provide a good
diagnostic tool for finding pseudobulges. Pseudobulges extend the
parameter correlations of photometric quantities ($r_e$, $\mu_e$,
$M_V$, and $n_b$). However in many of these parameters it is unclear
if pseudobulges actually show a correlation on their own.

We can compare the structural properties of the bulges in our sample
the output of simulation. Unfortunately simulations of galaxies that
resemble real galaxies including stars, dark matter, gas and star
formation (and possibly feedback) are quite difficult. Thus there does
not exist a statistically relevant set of simulations for full
comparison. Nonetheless, we can compare our results to those that
currently exist.

\cite{debattista2004} provides a set of simulations that generate
bulges from bar buckling in pure stellar (no gas) systems. The
resulting bulges typically have $n_b\sim1.5$, which is consistent with
what we find. The bulges in our sample tend to be more round than
their associated outer disk, and \cite{debattista2004} separate their
bulges based on flattening ratios. So we will only compare to those
bulges that are less round than the outer disk. Their simulated
galaxies have $B/D=0.2-1.0$ (where $B/D$ is bulge-to-disk light
ratio). Where as our sample has a median $B/D=0.12$ with a standard
deviation of $0.20$. Further, they are able to recover the coupling of
$r_e$ and $h$ that we find in pseudobulges. Thus our pseudobulges tend
to be a bit smaller. It may be that bar-buckling is one way to make a
pseudobulge, as indicated by the the fact that those simulated bulges
from \cite{debattista2004} are contained within the set of
pseudobulges, but do not span the whole range in properties.

We reiterate a statement by \cite{andredak1998}, that is also
discussed in \cite{kk04}, namely that we do not really understand why
pseudobulges should have a certain S\'ersic index. It is
understandable that mergers would drive up the S\'ersic index, as
discussed by \cite{vanalbada82} and \cite{kormendy2006virgo}, and thus
classical bulges are found with higher S\'ersic index.
\cite{aguerri2001} simulates minor mergers (accretion of satellites on
the order of the mass of the bulge) in bulge-disk galaxies. They show
that the S\'ersic index grows as the amount of mass accreted becomes
larger.  \cite{eliche2006} takes the study of satellite accretion to
lower densities than those simulations of \cite{aguerri2001}.  Their
simulations do not include gas and star formation, and thus as with
the \cite{debattista2004} simulations they should be read with that
caveat in mind. They find that low density satellite accretion does
not necessarily drive S\'ersic index above the critical value we
find($n_{crit}\sim2$). Yet, they also find that satellite accretion
leads to a simultaneous increase in $B/T$.

In Fig.~\ref{fig:btn}, we therefore plot the bulge-to-total ratio,
$B/T$, and the ratio of the bulge half-light radius to the scale
length of the disk, $r_e/h$, against $n_b$. There is a tight
correlation for classical bulges. However, pseudobulges do not follow
this correlation.  In fact, neither $B/T$ nor $r_e/h$ correlate with
$n_b$ in pseudobulges at all in the range of parameters shown,
spanning an order of magnitude in $B/T$. The absence of correlation
between $n_b$ and $B/T$ in pseudobulges seems to be pointing to a
non-merger driven formation scenario for pseudobulges. Yet, these
results are suggestive at best.  There may be some underlying
correlation that is destroyed by other factors (e.g.~gas fraction or
collision parameters). More work is needed from both simulations and
observations. That we do not see correlation of $n_b$ and $B/T$ for
low-S\'ersic index bulges (as found in \citealp{eliche2006}) may be
indicating that the classical bulges we observe today are the products
of multiple mergers, and possibly at higher redshifts there is a
population of low-S\'ersic index classical bulges.  Conversely it is
entirely possible that a population of small classical bulges exists,
yet they are embedded within pseudobulges. In this case the mass of a
bulge, and hence the $B/T$ within a specific galaxy would be coming
from multiple mechanisms.

If pseudobulges are built by secular evolution, the simplest mechanism
controlling the S\'ersic index in bulges is that $n_b$ grows with time
as the bulge-to-total ratio increases.  However, as discussed above,
it appears that there is not a strong connection between $B/T$ and
S\'ersic index.  This would imply that S\'ersic index in pseudobulges
is not a time-dependent quantity. We wish to emphasize, though, that
the error in the measurement of $n_b$ is large, and might be masking
an underlying weak correlation. Also, by focusing on intermediate-type
galaxies, where both pseudobulges and classical bulges occur, we miss
very low $B/T$ systems. Expanding a sample to later types might reveal
a weak correlation \citep{graham2001}.  Since $r_e$ correlates tightly
with $h$ for pseudobulges, it is no surprise that $r_e/h$ does not
correlate with $n_b$ in pseudobulges. The linear coupling of $r_e$ and
$h$ is well established in late type galaxies \citep{courteau1996}.

It may be that the S\'ersic index of pseudobulges is merely another
manifestation of the dynamical state of the system. Stars in bulges
with larger amounts of random motion often take on radial orbits, thus
climbing higher out of the potential well. Thus there is more light at
large radius, increasing $n_b$.  However, stars in orbits with higher
amounts of angular momentum would be less likely to take on radial
orbits and thus bulge light would contribute less at large radius;
driving $n_b$ down. The observation that the distribution of S\'ersic
indices in bulges of galaxies from E to Sc is bimodal then strengthens
the claim that what we are calling pseudobulges are not merely the
low-mass counterparts of the same phenomena that form classical bulges
and elliptical galaxies. This description fits well with the theory
that secular evolution forms pseudobulges, and mergers, whether by a
single event or succession of events, form classical bulges. The
higher angular momentum (and thus low S\'ersic index) systems have not
had major mergers and thus have not experienced the violent processes
that lower the ratio of rotational velocity to velocity
dispersion. What is left unknown is why pseudobulges are not
exponential, and also what keeps them from having larger S\'ersic
indices.

We conclude by returning to the results of Fig.~\ref{fig:bthist}, and
the implications of the range of $B/T$ for pseudobulges. It is now
well known that the presence of many low-$B/T$ systems presents a
problem for current $\Lambda$CDM galaxy formation theories
\citep{donghia2004}. In our sample we find no classical bulge galaxies
with $B/T<0.1$. Recent studies which compare merger histories in
$\Lambda$CDM simulations to the observed frequency of bulgeless
galaxies suggest that either there are too many mergers in the
simulations, or that disks must be much more robust to the merging
process than previously thought
\citep{stewart2007,koda2007}. Pseudobulges span a range of $B/T$ from
0.35 to zero. If pseudobulges form through internal-disk processes
then a galaxy with a pseudobulge can be thought of as a pure disk
galaxy. Thus current estimates of the number of low-$B/T$ systems
could only be thought of as lower-limits; the existence of
pseudobulges would make the problem of forming bulgeless systems even
more pressing.

\acknowledgments

We would like to thank Prof.\ John Kormendy \& Dr.\ Mark Cornell for
countless discussions on this work. Also, DBF would like to thank
Prof.\ Alex Filippenko and University of California at Berkeley for
providing him with a place to do much of this research.

DBF acknowledges support by the National Science Foundation under
grant AST 06-07490. This work is based in part on observations made
with the Spitzer Space Telescope, operated by the Jet Propulsion
Laboratory, California Institute of Technology under a contract with
NASA. Support for this work was provided by NASA through an award
issued by JPL/Caltech.

This paper makes use of data obtained from the Isaac Newton Group
Archive which is maintained as part of the CASU Astronomical Data
Centre at the Institute of Astronomy, Cambridge. This research has
made use of the NASA/IPAC Extragalactic Database (NED) which is
operated by the Jet Propulsion Laboratory, California Institute of
Technology, under contract with the National Aeronautics and Space
Administration. This research has made use of the NASA/ IPAC Infrared
Science Archive, which is operated by the Jet Propulsion Laboratory,
California Institute of Technology, under contract with the National
Aeronautics and Space Administration. Some of the data presented in
this paper were obtained from the Multi-mission Archive at the Space
Telescope Science Institute (MAST). STScI is operated by the
Association of Universities for Research in Astronomy, Inc., under
NASA contract NAS5-26555. Support for MAST for non-HST data is
provided by the NASA Office of Space Science via grant NAG5-7584 and
by other grants and contracts. Funding for the SDSS and SDSS-II has
been provided by the Alfred P.\ Sloan Foundation, the Participating
Institutions, the National Science Foundation, the U.S.\ Department of
Energy, the National Aeronautics and Space Administration, the
Japanese Monbukagakusho, the Max Planck Society, and the Higher
Education Funding Council for England. The SDSS Web Site is {\tt
  http://www.sdss.org/}. The SDSS is managed by the Astrophysical
Research Consortium for the Participating Institutions. The
Participating Institutions are the American Museum of Natural History,
Astrophysical Institute Potsdam, University of Basel, Cambridge
University, Case Western Reserve University, University of Chicago,
Drexel University, Fermilab, the Institute for Advanced Study, the
Japan Participation Group, Johns Hopkins University, the Joint
Institute for Nuclear Astrophysics, the Kavli Institute for Particle
Astrophysics and Cosmology, the Korean Scientist Group, the Chinese
Academy of Sciences (LAMOST), Los Alamos National Laboratory, the
Max-Planck-Institute for Astronomy (MPIA), the Max-Planck-Institute
for Astrophysics (MPA), New Mexico State University, Ohio State
University, University of Pittsburgh, University of Portsmouth,
Princeton University, the United States Naval Observatory, and the
University of Washington. This publication makes use of data products
from the Two Micron All Sky Survey, which is a joint project of the
University of Massachusetts and the Infrared Processing and Analysis
Center/California Institute of Technology, funded by the National
Aeronautics and Space Administration and the National Science
Foundation.


\end{document}